 \documentclass[authoryear,preprint,review,12pt]{elsarticle}

\usepackage{amssymb}
\usepackage{bm}

\usepackage{caption}
\usepackage{subfig}

\journal{Ocean Engineering}

\begin{document}

\begin{frontmatter}



\title{An experimental study of using artificial reefs as scour protection around an offshore wind monopile}


\author[S1,S2]{Xin Liu}
\author[S1,S2]{Jianjun Chen}
\author[S1,S2]{Yu Lei}
\author[S1,S2]{Ruichao Liu}
\author[S1,S2]{Yiming Zhou}
\author[J2]{Han Li\corref{cor1}}
\author[J2]{Jing Yuan}
\cortext[cor1]{Corresponding author, email: lihan24@mails.tsinghua.edu.cn}

\affiliation[S1]{organization={China Huaneng Clean Energy Research Institute},
            city={Beijing},
            postcode={102209}, 
            country={China}}
            
\affiliation[S2]{organization={National Energy R\&D Center of Offshore Wind Power Engineering and Operation},
            city={Beijing},
            postcode={102209}, 
            country={China}}
            
\affiliation[J2]{organization={Department of Hydraulic Engineering, Tsinghua Univesity},
            city={Beijing},
            postcode={100084}, 
            country={China}}

\begin{abstract}
Artificial reefs (ARs) are man-made structures deployed on the seabed to support benthic marine ecosystems. Their presence significantly damps the local flow and therefore can be used for scour protection of offshore wind monopiles. Although the concept appears feasible, the underlying flow-sediment process is very complex and has yet been systematically investigated. To fill in this gap, a set of fixed-bed flume tests were conducted to reveal the hydrodynamic details of two typical AR shapes (cubic and hemisphere) tightly placed around a monopile in a 3$\times$3 pattern. In parallel, a set of live-bed tests were conducted to demonstrate the AR's efficiency in scour protection. The cubic ARs almost eliminate the downward flow on the upstream side of the monopile and reduces the wake flow by 50$\sim$80\%. The hemisphere ARs also significantly weaken the wake flow but guide descending flow in front of the monopile. While cubic ARs decrease upstream and downstream scour depth by up to 100\%, their edge scour can lead to ARs displacement and hence reduce scour protection. The hemisphere ARs provided less scour reduction, but also less edge scour, making them more adaptive to morphology changes. Based on these findings, an optimized AR layout was proposed.
\end{abstract}

\begin{graphicalabstract}
\end{graphicalabstract}

\begin{highlights}
\item Combine fixed- and live-bed tests to study scour of monopile surrounded by artificial reefs.
\item ARs can significantly reduce local flow and prevent scour by 30\% to 100\%. 
\item Edge scour of ARs can have negative effects.
\end{highlights}


\end{frontmatter}

\section{Introduction}

Over the past decades, the growing demand for renewable energy has driven the rapid development of offshore wind farms (OWFs) over the world \citep{RN9}. A main challenge for OWF, commonly using monopile foundations, is the local scour caused by marine currents or waves, which severely affects the bearing capacity and structural dynamics \citep{RN16}. Scour protection is the essential solution to this problem and has been studied extensively by many researchers. Some common types of scour protections include ripraps, sand bags, soil consolidation, etc. Artificial reefs (ARs), which are man-made structures placed on the seabed to restore marine ecosystems, are widely used in coastal areas \citep{RN31}. Their size is of order O(1 m) to O(10 m), which is comparable to the size of ripraps or monopiles, and they usually use low-cost and environmentally materials, concrete for instance, so they are considered as an innovative solution for scour protection \citep{RN19}.

\begin{figure}
  \centering
  \includegraphics[width=\linewidth]{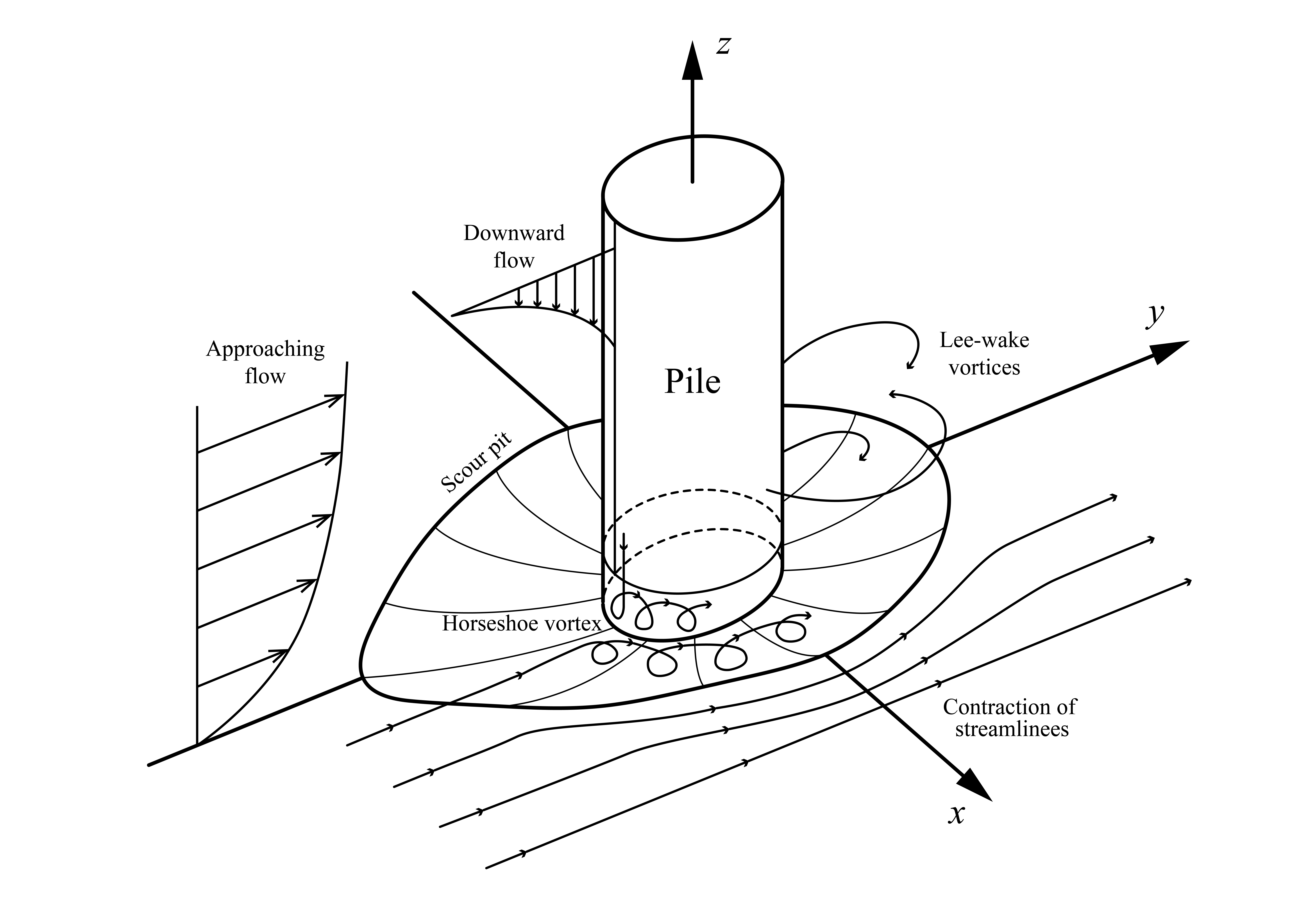}
  \caption{Flow characteristics and scour phenomenon around a monopile.}
  \label{fig_2}
\end{figure}

The mechanics of scour around monopile in the marine environment have been thoroughly summarized by \citet{RN5} and \citet{RN7}. Fig.~\ref{fig_2} illustrates the main hydrodynamic process for near-bed current flows passing a monopile. First, the presence of the monopile forces the water in front of the pile to flow downward, which finally impacts the seabed and rolls up to form a horseshoe vortex. Second, the streamlines passing the lateral sides of the pile are squeezed and contracted, leading to a local flow acceleration. Third, the wake vortex shed from the pile is generated on the downstream side. When bottom shear stress is amplified by these processes, sediment incipient motion and transport will occur. Besides, numerous studies have been conducted on the prediction of scour development and factors determining the scour results using flume experiments under current-only \citep[e.g.,][]{RN52,RN41,RN53,RN3}, wave-only \citep[e.g.,][]{RN21,RN17} or combined current-wave \citep[e.g.,][]{RN14,RN39,RN24,WANG} conditions. In addition, some high fidelity numerical simulations have been reported in the literature \citep[e.g.,][]{RN50,RN54,RN51}, but the high computational cost limits the role of numerical simulation in studying scours \citep{RN4}.

Various scour protections have been proposed, which can be divided into two groups \citep{RN18}: to reduce the energy of water flow and to enhance the stability of bed sediment. The former includes adding collars to the bottom of the monopile \citep[e.g.,][]{RN11} or installing bionic grass on the sea bed \citep[e.g.,][]{RN46}. These extra structures alter the local flow conditions and prevent the seabed from direct flow impact. The latter changes the physical properties of the seabed, such as soil consolidation \citep[e.g.,][]{RN15}, or covering the seabed with an armor layer. Riprap protection, which covers the seabed with filter layers and large armor stones, is the most widely adopted measure in practice. \citet{RN33} developed an effective Opti-Pile Design Tool to predict scour depth, scour width, and required rock size based on data from flume experiments. To explain the edge scour of riprap protection, \citet{RN32} resolved the three-dimensional flow field around a monopile through particle image velocimetry (PIV) and bed shear stress measurements. They found a series of vortices emerging on the bed surface, which play a governing role in protection failure. Using flume experiments, \citet[][\citeyear{RN47}]{RN6} proposed several empirical criteria for engineering design. Stone displacements and topographic changes were measured under various flow conditions to derive equations of required rock size for riprap protection. These comprehensive research efforts provide a solid basis for applying ripraps for scour protection in offshore wind farms.

\begin{figure}
  \centering
  \includegraphics[width=0.5\linewidth]{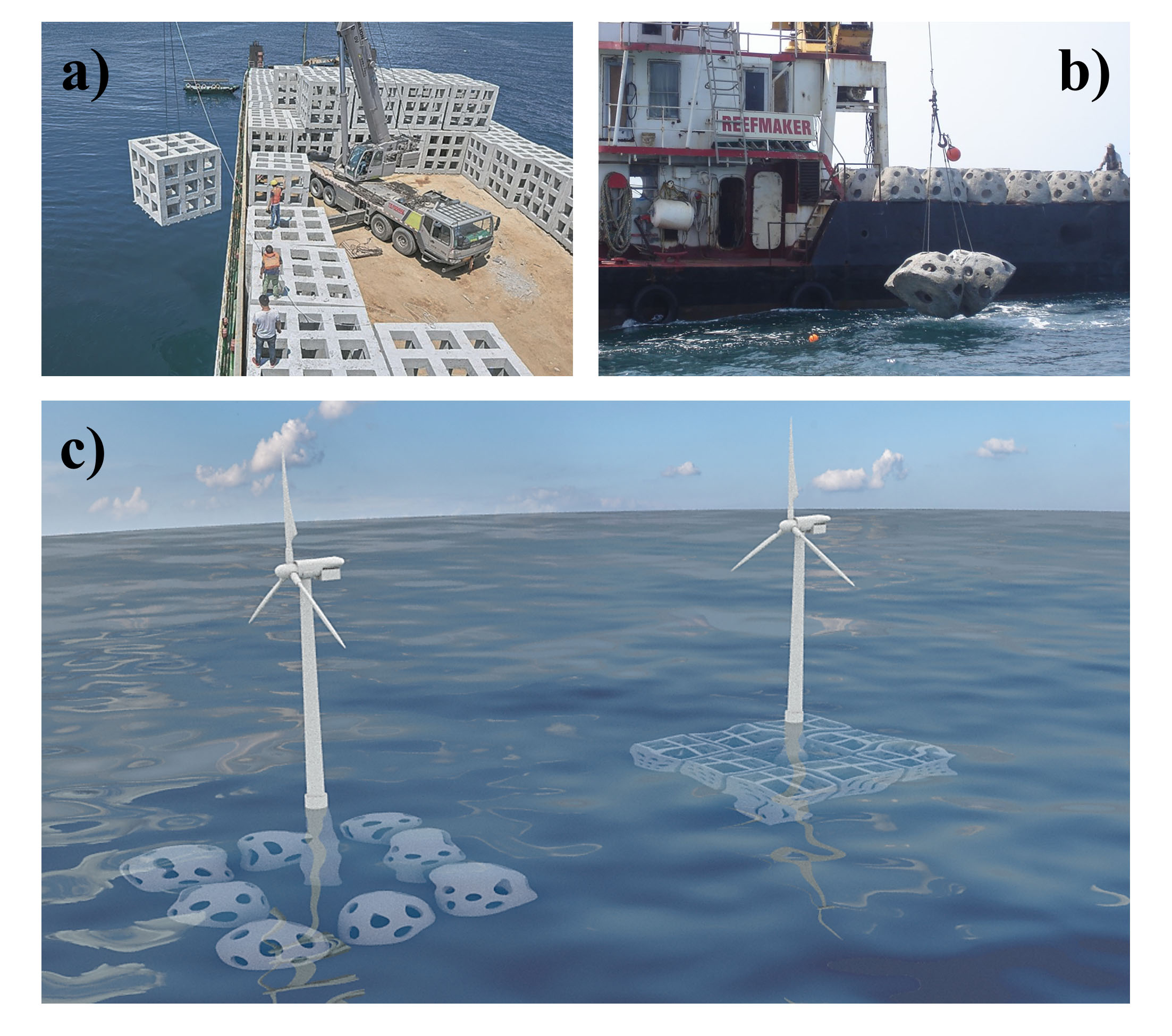}
  \caption{The concept of using ARs as scour protection around monopile foundations: a) Cubic reefs applied in Fengjia Bay, Hainan, China; b) Reef balls deployed in Deadman's Bay, Florida, USA; c) A conceptual plan on combination of marine ranching and offshore wind farm.}
  \label{fig_1}
\end{figure}

Compared to conventional scour protection measures, using ARs have some advantages. First, ARs significantly contribute to the ecological sustainability of wind farms. Existing ARs for large-scale applications are predominantly built with concrete for its low cost and environmental friendliness \citep{RN45}. It is understood that scour protection structures, even ripraps, can increase the abundance of marine species \citep{RN19}. Well-designed ARs can provide larger and more complex habitats than ripraps to support a diverse marine community \citep{RN44}. The flows in and around ARs can be guided to create suitable local hydrodynamic environments for various species. Second, optimizing the structure of ARs can better dissipate the energy of water flow, thus improving their efficiency of scour protection. Nowadays, typical shapes of AR include blocks (Fig.~\ref{fig_1}a), reef balls (Fig.~\ref{fig_1}b) and pipes \citep{RN22}. Extensive research has investigated the influence of AR structures on the local flow features \citep[e.g.,][]{RN49,RN34,RN28}. It is possible to form a scour protection by arranging ARs around a monopile, as illustrated in Fig.~\ref{fig_1}c. However, there is little research showing whether this concept works. \citet{RN20} established a numerical model to compare the performance of ARs for monopile scour protection under different AR layouts. Their results indicate that arranging ARs along the main stream direction effectively reduces scour, but arranging ARs perpendicular to the main stream may have a negative impact.

In order to obtain insights on whether and how ARs can prevent scour around monopile in current flows, two typical AR structures were tested through both fixed-bed tests, which aim at understanding the flow processes, and live-bed tests, which demonstrate the performance of ARs for reducing scour. The experimental results of the two groups were examined together to yield findings that can serve as the basis for practical applications. The remaining parts of the paper are organized as follows. Section 2 describes the experimental setup. Section 3 focuses on the flow field characteristic derived from fixed-bed tests, while Section 4 discusses the scour phenomena in live-bed tests. Also in Section 4, a suggested layout of AR is proposed based on our findings. Finally, the conclusions are given in Section 5.

\section{Experimental setup}

This study included two groups of tests: (1) fix-bed tests, which give flow measurements at a few 2-dimensional-vertical cuts of the ARs, and (2) live-bed tests, in which equilibrium scour depth and width were measured. The two groups were conducted in separate flumes in the Hydraulics Laboratory at Tsinghua University. This section first introduce the selected ARs, followed by the presentations of the two flumes and associated test procedures.

\subsection{Physical models of ARs}

\begin{figure}
  \centering
  \includegraphics[width=0.49\linewidth]{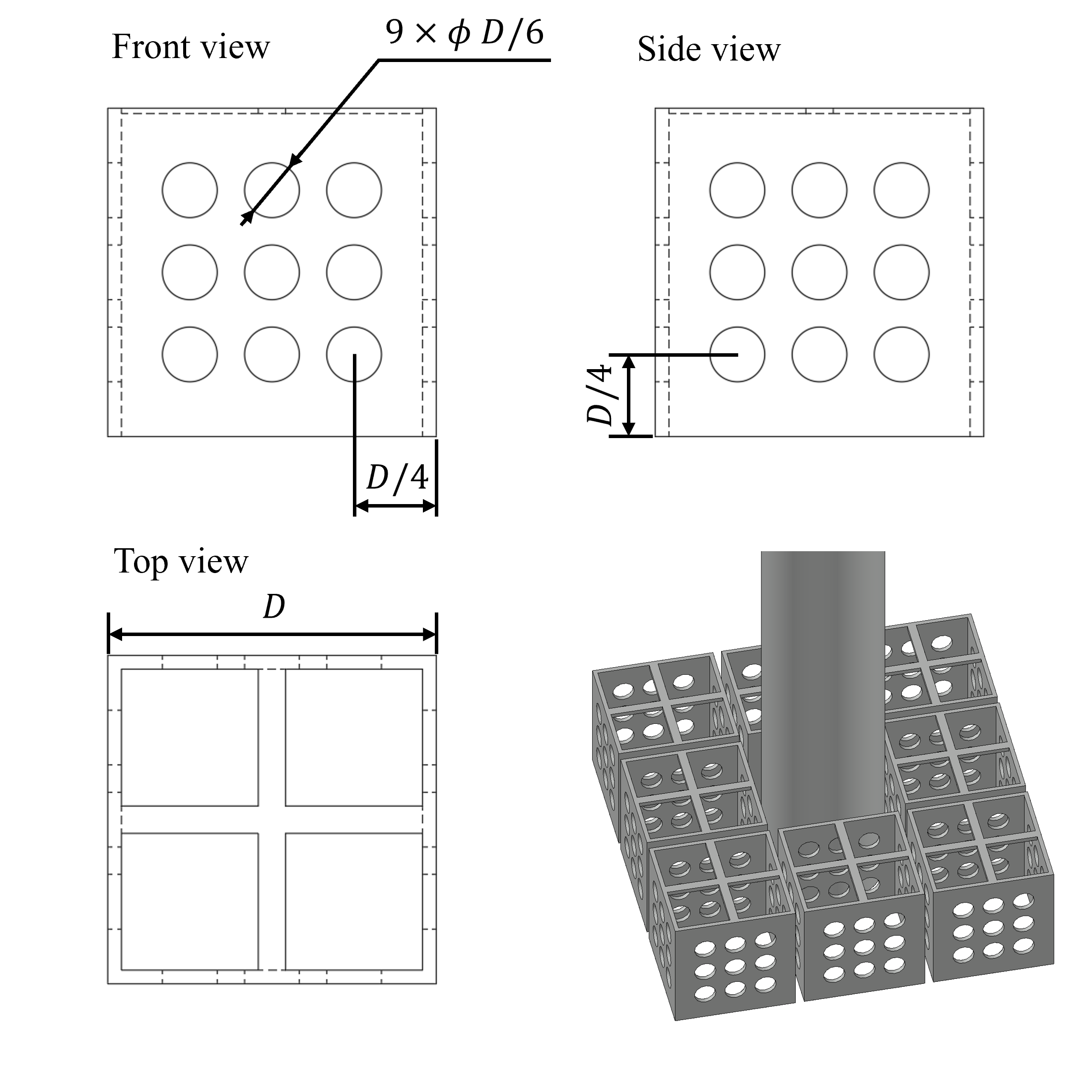}
  \includegraphics[width=0.49\linewidth]{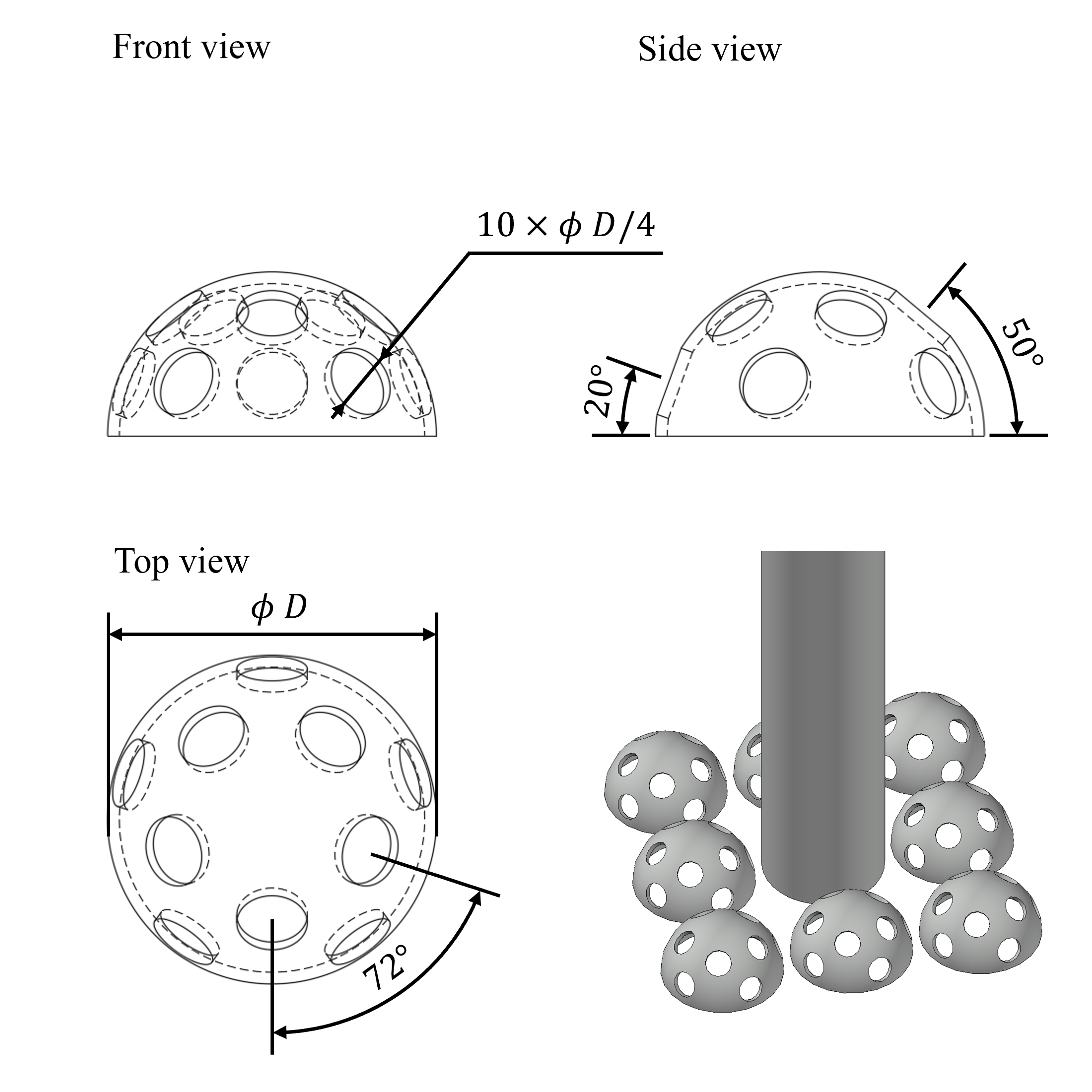}
  \caption{Structures and arrangement methods of ARs.}
  \label{fig_4}
\end{figure}

The two designs of ARs, i.e., Cubic AR (C-AR) and Hemisphere AR (H-AR), selected in this study are shown in Fig.~\ref{fig_4}. Both designs are typically used in marine ranch constructions for their unique ecological functions (see Fig.~\ref{fig_2}). The C-AR has a cubic shape with 3$\times$3 holes on the four lateral sides. The diameter of each hole is 1/6$D$, where $D$ is the edge length of the cubic. Thus, the porosity of the surface is 19.63\%. The top is mostly open, except for two cross bars for ensuring the integrity of the AR. The bottom is completely open. With this design, the flow is allowed to enter the C-AR from the top and holes on four sides, and the seabed underneath the C-AR is exposed to the internal flow. The H-AR has ten holes on its surface, which are uniformed spaced. The diameter of the hole is 1/4$D$, where $D$ represents the diameter of the hemisphere. Therefore, the porosity of its surface is 31.25\%. The bottom is also open. Both ARs have large internal void space. It is possible to add some secondary structures inside the ARs for creating suitable spaces for certain marine species. The main objective of this study is to understand how the overall shape of the ARs affects the flow around a monopile, so we chose to keep the ARs empty. Future studies are expected to understand the effects of modifying the internal space. The C-AR represents AR design with sharp edges or corners, which may significantly block the local flow. The H-AR represents AR design with streamlined design, which has less impact on local flow.

In this study, the dimension of ARs is the same as the diameter of the monopile, which allows placing ARs tightly around the monopile in a 3$\times$3 array, as shown in Fig.~\ref{fig_4}. This is the tightest layout, so the ARs can presumably best reduce the flows in the vicinity of the monopile and protect it from scour. Although this layout may not be the optimal, e.g., increasing the spacing among ARs may not necessarily promote scour, it serves as the basis for further studying other ways of placing ARs.

The live- and fixed-bed tests were conducted in flumes with different dimensions, and the aims of these tests were different. Thus, two sets of AR models were produced. In the fixed-bed tests, it is necessary to measure the flow inside the ARs, so the models were made of transparent acrylic (Fig.~\ref{fig_3}) to facilitate the measurements of PIV. The dimension of AR is 14 cm, which is the same as the model pile’s diameter. The thickness of the ARs shell is 0.5 cm. For live-bed tests, the AR models were made of aluminum alloy (Fig.~\ref{fig_5}), which has a density of $\rho$ = 2.75 g/cm$^3$. This density is close to that of concrete, so the model ARs can mimic the behavior of concrete ARs in the field. The dimension of these ARs and the diameter of the model pile are both 5 cm, which are smaller than that of ARs in the fixed-bed test. This is because the live-bed tests were conducted in a smaller flume. The model must be smaller to avoid the side-wall effects of the flume.

In prototype conditions, the diameter of the monopile foundation can be 2$\sim$5 m for small to middle-size OWFs, and ARs can be made with the same size. Thus, the geometry scale of our models is 1:15$\sim$40 and 1:40$\sim$100 for live- and fixed-bed tests, respectively.

\subsection{Fixed-bed test}

\begin{figure}
  \centering
    \includegraphics[width=\linewidth]{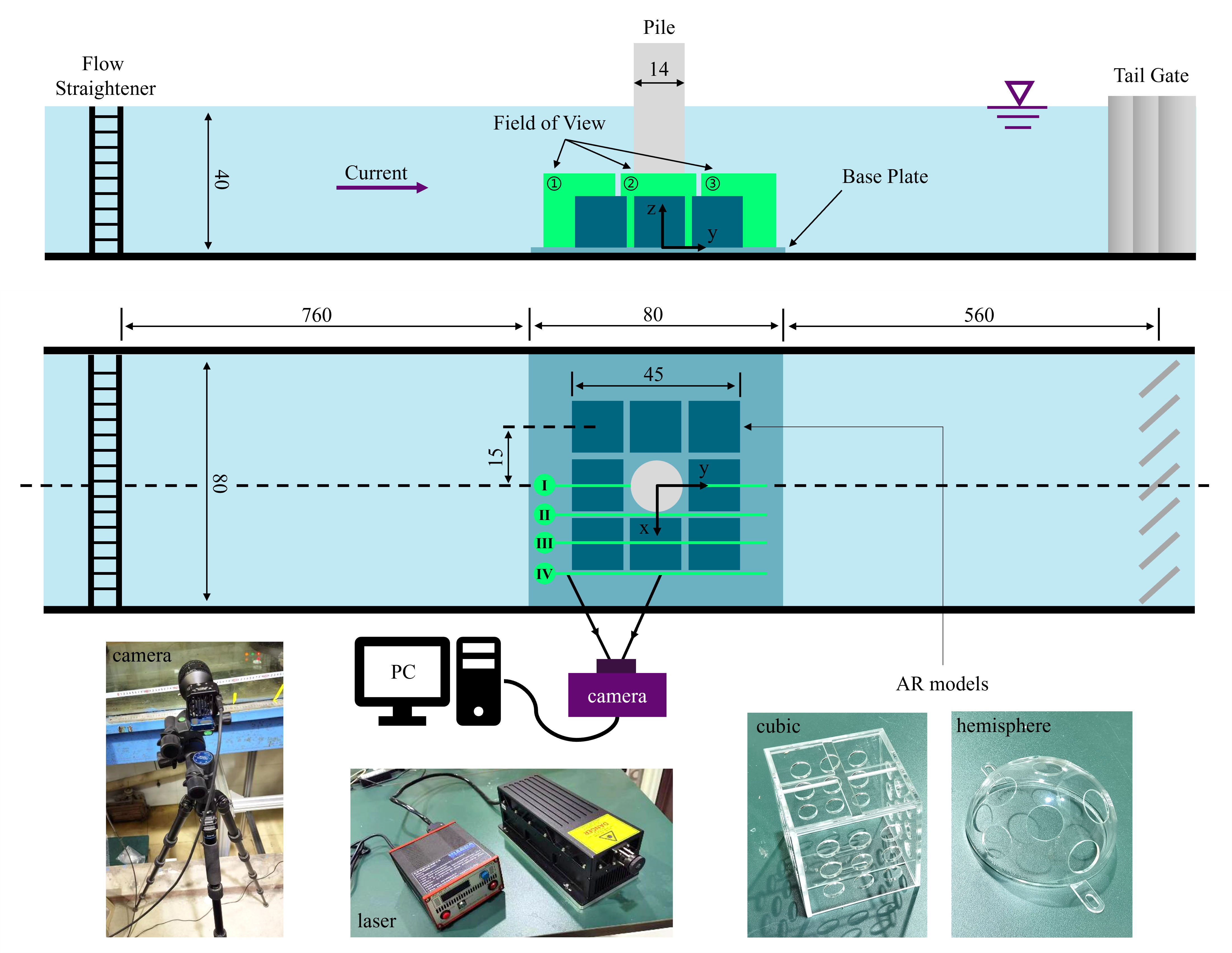}
  \caption{Experimental setup for fixed-bed tests in top view and side view (Dimensions in centimeter, not to scale).}
  \label{fig_3}
\end{figure}

The fixed-bed tests were conducted in a current flume, which is 14 m in length, 0.8 m in width, and 0.8 m in height, as shown in Fig.~\ref{fig_3}. The steady current in the flume is driven by a frequency-controlled centrifugal pump, whose flow rate can be up to 0.2 m$^3$/s. The flume’s bottom and side walls are all made of glass to facilitate PIV observations. A PVC pipe of $D_1$ = 14 cm diameter, which depicts a common monopile applied in offshore wind farms, is surrounded by cubic ARs of 14 cm size or hemisphere ARs of 14 cm diameter. All AR units are evenly spaced with a 1 cm gap and bolted on a transparent breadboard base. Have such a narrow gap is because it is impossible to avoid gaps when installing ARs in the field.

The PIV measurements were obtained using a high-speed camera (IDT Vision, XSM-3520, resolution: 2560$\times$1440 pixels, sampling rate: up to 2350 fps) with a Nikon lens and a continuous laser source (VIASHO, VA-II-N-532nm-20W). Seeding particles with a diameter of 10 $\mu$m were added to the flume and illuminated by the laser light from below the flume. The laser sheet was aligned in the streamwise direction and positioned from the center of the flume to the edge of the models, as the planes I to IV labeled in Fig.~\ref{fig_3}. Plane I covers the centerline cut of the flume, which allows us to understand the upstream descending flow and downstream wake flow of the pile. Plane II is placed within the 1 cm gap between two rows of ARs, which allows us to see whether the flow is squeezed into these gaps and forms a high-speed jet. Plane III 
is a centerline cut of a streamwise row of ARs, which allows us to see the internal flow fields. Plane IV is along the outer streamwise edge of the AR arrays, so the observed flow can be used to interpret edge scour effects. Limited by the width of laser sheet, each plane is further separated into three fields of view (FOVs, see 1 to 3 in Fig.~\ref{fig_3}) except for plane I, which has two FOVs in the upstream and downstream sides of the pile’s centerline cut. For each FOV, a collection of 1000 PIV pictures was captured at a frequency of 250 Hz and an exposure time of 236 $\mu$s.

This research takes on typical offshore wind farms in China as the background, which have water depth of approximately 5$\sim$20 m and maximum tidal current between 1.0$\sim$2.0 m/s. Using Froude-number scaling, the model-scale flow velocity is about 0.2$\sim$0.4 m/s and the water depth is about 20 to 80 cm, if a geometry scale of 1:25 is assumed. Thus, the following flow condition of fixed-bed tests was selected: the sectional mean flow velocity, $u_0$ = 0.30 m/s, and water depth $h$ = 40 cm, as shown in Table.~\ref{tab_fixed-conditions}. During each test, we first adjusted the pump power and the tail gate opening until the desired flow rate and water depth were achieved. Once the flow conditions were set, the laser source, mounted on a metal frame with the ability to move freely, was positioned at the desired locations to illuminate a FOV. Next, the camera position and focal length were adjusted to capture the entire FOV clearly. After these preparations, PIV measurements were conducted, and the above process was repeated for all the FOVs.

\begin{table}
  \centering
  \caption{Fixed-bed experiment test conditions.}
  \begin{tabular}{ccccccc}
    \hline
    Test & AR structure & $h$ [m]   & $Q$ [m$^3$/h] & $u_0$ [m/s] \\
    \hline
    F1 & control  & 0.40   &  346.1  & 0.30  \\
    F2 & cubic  & 0.40   &  346.7  & 0.30  \\
    F3 & hemisphere  & 0.40   &  344.9  & 0.30 \\  
    \hline
  \end{tabular}
  \label{tab_fixed-conditions}
\end{table}

\subsection{Live-bed test}

\begin{figure}
  \centering
    \includegraphics[width=0.7\linewidth]{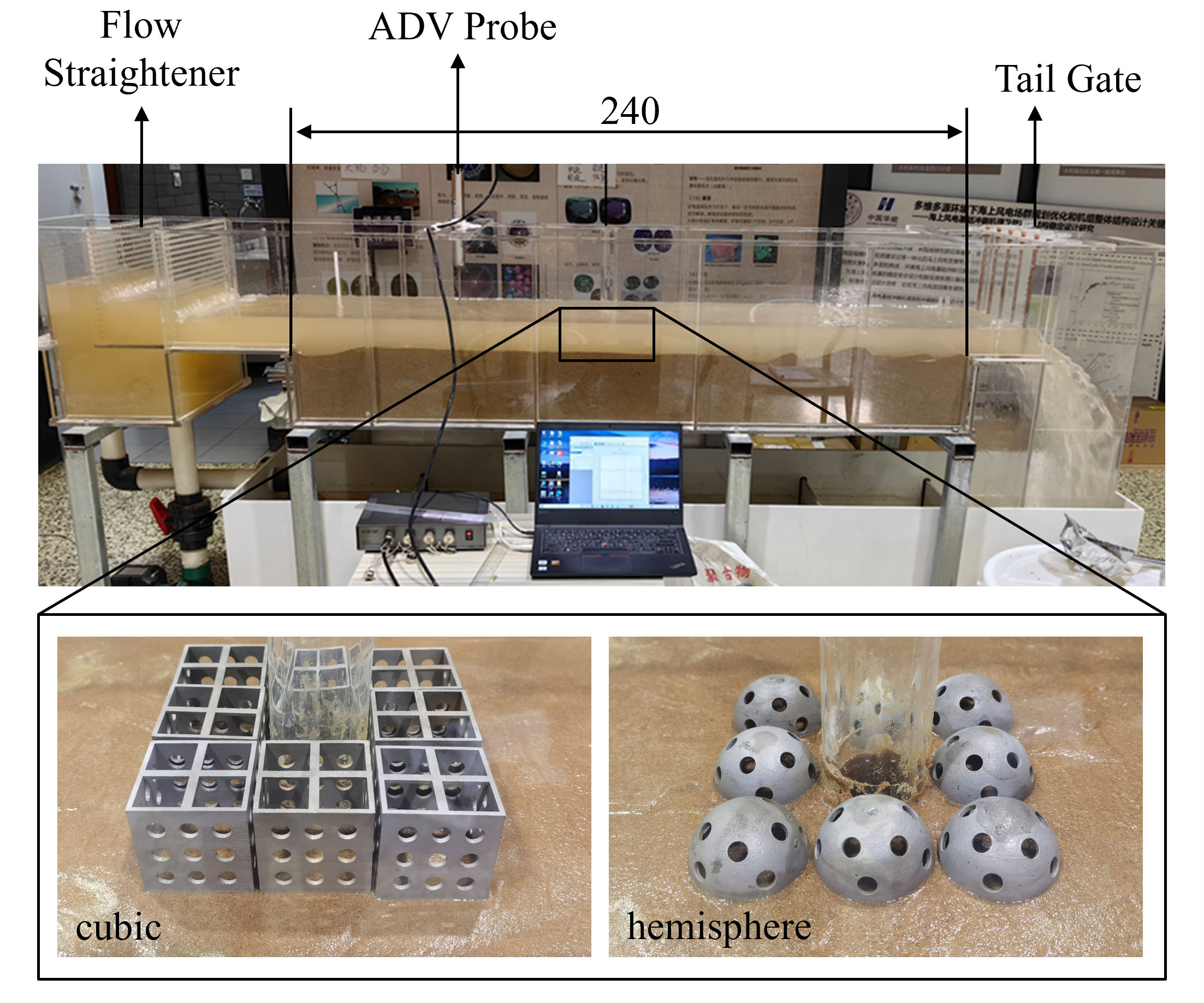}
  \caption{Test flume for scour measurements in side view. Dimensions in centimeter.}
  \label{fig_5}
\end{figure}

Live-bed tests were conducted in a smaller current flume, which is 0.5 m wide, 2.4 m long and 0.6 m deep, as shown in Fig.~\ref{fig_5}. Its test section has a 20 cm-deep sediment pit filled with sand characterized by a median diameter $d_{50}$ = 0.235 mm, a geometric standard deviation $\sigma_g$ = $\sqrt{d_{84} / d_{16}}$ = 1.6 and specific density $s$ = 2.65. The flume can generate steady flow with maximum flow rate of 0.01 m$^3$/s. An Acoustic Doppler Velocimetry (ADV) probe was installed approximately 1 m downstream from the entrance to monitor the income flow speed. The model pile of $D_2$ = 5 cm was fixed and extended vertically downward into the sand bed. The AR units were placed carefully around the pile following the same model setup in the fixed-bed tests.

Due to the smaller size of the flume, the geometric scale of live-bed tests is about 1:40 to 1:100. We surveyed 11 previous experimental studies of monopile-foundation scour, and found that the geometric scale is mainly between 25 to 100, so our tests’ scale, although is within the lower end, can yield acceptable results. We, however, acknowledge that more tests with larger geometric scale is required to further verify the quantitative results of our tests. For scour study, a key parameter is the Shields parameter, $\theta$, which directly characterizes sediment transport processes \citep{RN8}. For many field conditions in China’s coastal area, $\theta$ is around the threshold value for incipient motion, $\theta_{cr}$, so we selected three mean flow velocities, i.e., 0.20 m/s, 0.25 m/s, and 0.30 m/s, which yield $\theta$ = 0.027, 0.043, and 0.062, respectively. These values of $\theta$ are smaller than, equal to, and larger than $\theta_{cr}$ = 0.043 for the sand used in our tests, so these flow conditions cover both clean-bed and live-bed scours. 

The complete test program is given in Table.~\ref{tab_live-conditions}. Tests L1 to L9 have the same AR layouts as in the fixed-bed tests. Test L10 to L14 have modified layouts based on findings from L1 to L9, which will be introduced later. The flow velocities $u_c$ at the sectional midpoint were obtained by the ADV. The duration of the test was mostly set at 1 hour. We have selected a few tests and increased their duration to 2 hours. By comparing the scour results after 1 hour and 2 hours, a negligible difference was observed, suggesting that 1 hour is sufficiently long to allow the development of scour pit. The general procedure of a live-bed test was as follows:

\begin{itemize}
  \item Step 1: Scrap the sand bed flat and carefully deploy the ARs around the monopile;
  \item Step 2: Fill the flume to the target water level and gradually accelerate the flow to the desired velocity;
  \item Step 3: Turn off the pump after running a test for 1 h, and wait for another 0.5 h to deposit the suspended sand;
  \item Step 4: Measure the scour geometry.
\end{itemize}

\begin{table}
  \centering
  \caption{Live-bed test conditions.}
  \begin{tabular}{cccccc}
    \hline
    Test & AR structure  & $h$ [m]  & $u_c$ [m/s] & $\theta=\frac{\tau}{(\rho_s-\rho)gD}$ [-] \\
    \hline
    L1 & control  & 0.10   &  0.20   &  0.027  \\
    L2 & control  & 0.10   &  0.25   &  0.043 \\
    L3 & control  & 0.10   &  0.30   &  0.062 \\
    \hline
    L4 & cubic  & 0.10   &  0.20   &  0.027  \\
    L5 & cubic  & 0.10   &  0.25   &  0.043  \\
    L6 & cubic  & 0.10   &  0.30   &  0.062  \\
    L7 & hemisphere  & 0.10   &  0.20  &  0.027    \\
    L8 & hemisphere  & 0.10   &  0.25  &  0.043   \\
    L9 & hemisphere  & 0.10   &  0.30  &  0.062   \\
    \hline
    L10 & cubic  & 0.10   &  0.30   &   0.062  \\
    L11 & hemisphere  & 0.10   &  0.30    &  0.062   \\
    L12 & cubic  & 0.10   &  0.30   &  0.062 \\
    L13 & hemisphere  & 0.10   &  0.30   &  0.062 \\
    L14 & cubic \& hemisphere & 0.10   &  0.30   &  0.062 \\
    \hline
  \end{tabular}
  \label{tab_live-conditions}
\end{table}

\section{Fixed-bed tests}

\subsection{Data analysis}
The PIV measurements were post-processed by PIVlab v3.00 \citep{RN44} based on Matlab. Out of 1000 total images at each FOV, the first two images in every set of 10 were taken as a pair. In other words, flow fields were captured at a frequency of 25 Hz for a duration of 4 s. The calibration was accomplished using the length of monopile and AR models. Background image of mean intensity was first subtracted from the original images to increase signal-to-noise ratio. Then a three-pass cross-correlation algorithm (FFT-based, final interrogation window size of 32 $\times$ 32 pixels with a 50\% overlap) was applied to obtain instantaneous velocity fields. Subsequently, the vector maps were filtered by a standard deviation filter with a threshold of 8 and a local median filter with a threshold of 3. Typical vector removal rates ranged from 5$\sim$15\% for two main reasons. First, refraction of laser light (when travelling obliquely through the transparent models) leaves some dark slices in the PIV image, which were filled with interpolated vectors. Second, some FOVs have large out-of-plane flow velocity, due to the 3-dimensionality of local flow, which were marked by white masks in the results.

To understand the characteristics of the flow field, dimensionless time-averaged velocity $\bm{\bar{u}}/u_0 $ and unsteadiness of velocity $ \Delta u / u_0 $ are given by:
\begin{equation}
    \bm{\bar{u}} / u_0 = \frac{\sum_{k=1}^{n} \bm{u_k}}{n} / u_0 ,
\end{equation}
and
\begin{equation}
    \Delta u / u_0 = \sqrt{ \frac{\sum_{k=1}^{n} \| \bm{u_k} - \bm{\bar{u}}\|^2 }{n-1} } / u_0 ,
\end{equation}\label{eq_flow_field}
where $n$ is the total number of the PIV frames, $\bm{u_k}$ is the velocity vector at a specific point in the frame $k$ and $u_0$ is the sectional mean flow velocity ($u_0$ = 0.30 m/s).

\subsection{The control test}

\begin{figure}
  \centering
    \includegraphics[width=\linewidth]{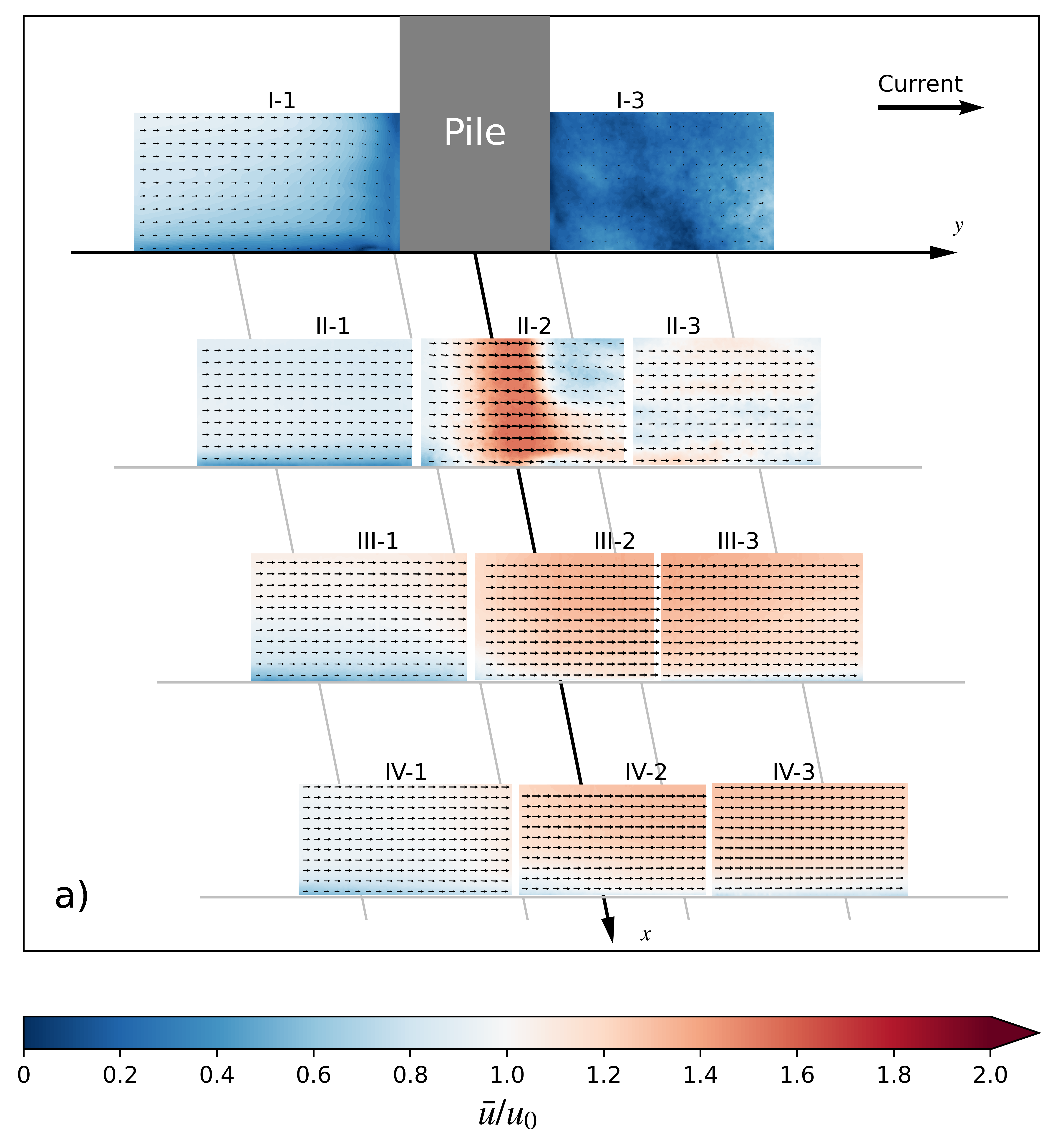}
\end{figure}
\begin{figure}
    \centering
    \includegraphics[width=\linewidth]{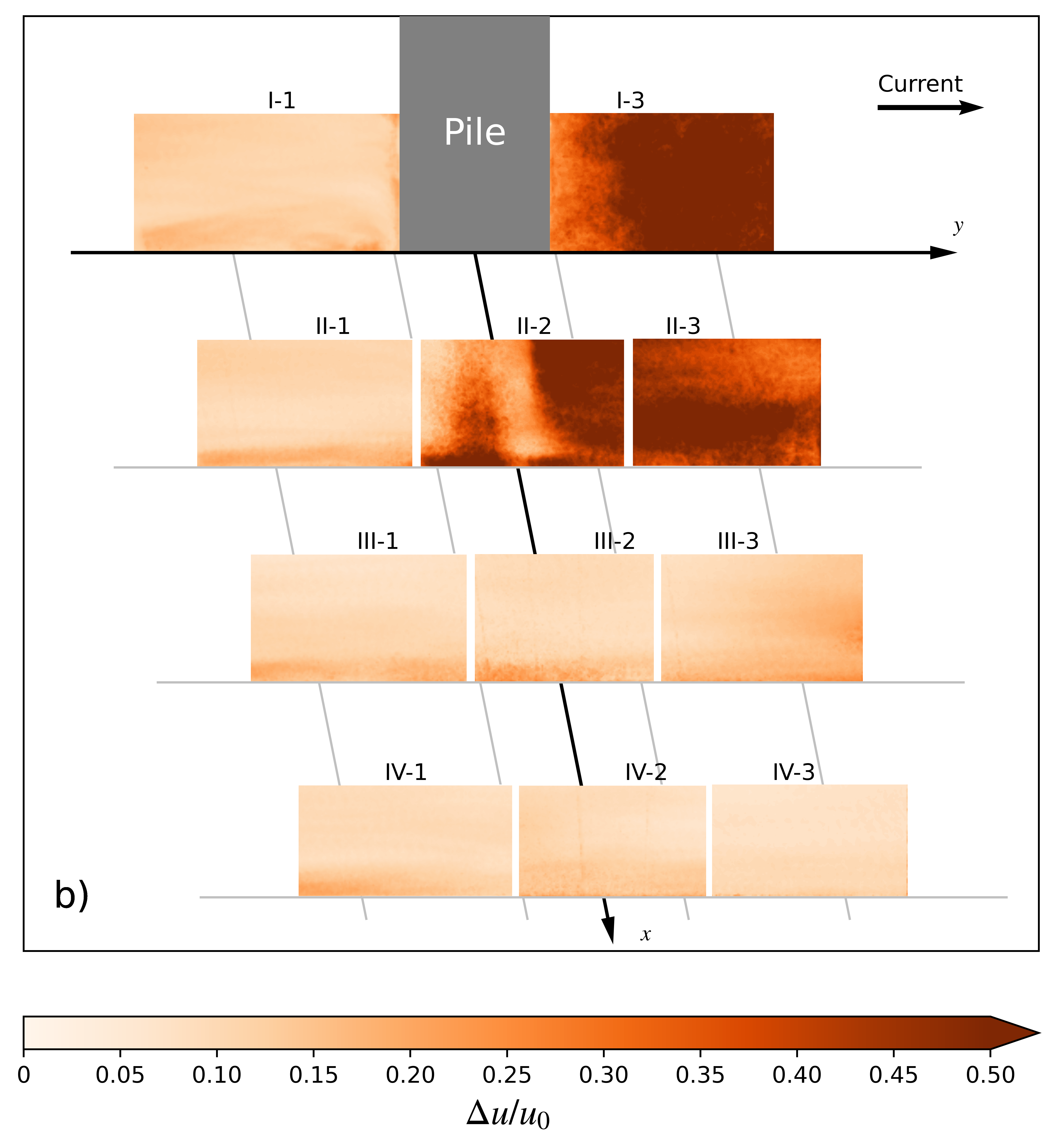}
  \caption{PIV measurement for test F1: (a) mean flow and (b) unsteadiness.}
  \label{fig_9_control}
\end{figure}

The observed flow in the control test shows the features of steady flow over a monopile in Fig.~\ref{fig_2}. Some key findings are highlighted as follows.

Panel I: An obvious reduction of $\bar{u}$ and a downward deflection of flow can be observed immediately upstream from the pile (FOV I-1 of Fig.~\ref{fig_9_control}), which represents the downward flow highlighted in Fig.~\ref{fig_2}. FOV I-3 captures the wake region behind the pile, characterized by a low averaged velocity and strong unsteadiness. This unsteadiness is primarily caused by flow separation after the pile and the vortex shedding.
 
Plane II: FOV II-2 covers the region where flow passes the lateral side of the pile. Here, the main incoming flow is forced to go around the pile, leading to an increased average velocity at about 1.5$u_0$. Another important feature observed in FOV II-2 is the enhanced unsteadiness near the bottom, which corresponds to the location of the horseshoe vortex. In FOV II-3, low $\bar{u}$ and high $\Delta u$ suggest that the turbulent wake extended into this region. 

Panel III and IV: These two panels are 15 cm and 22.5 cm away from the centerline and exhibit similar flow patterns. In FOV III-3 and IV-3, $\bar{u}$ exceeds the incoming flow, which suggests the wake region with low $\bar{u}$ is confined within $\pm$15 cm from the centerline. The low levels of unsteadiness in both III-3 and IV-3 further confirm that these two regions lie outside the pile's wake zone.

\subsection{Flows with C-ARs}
\begin{figure}
  \centering
    \includegraphics[width=\linewidth]{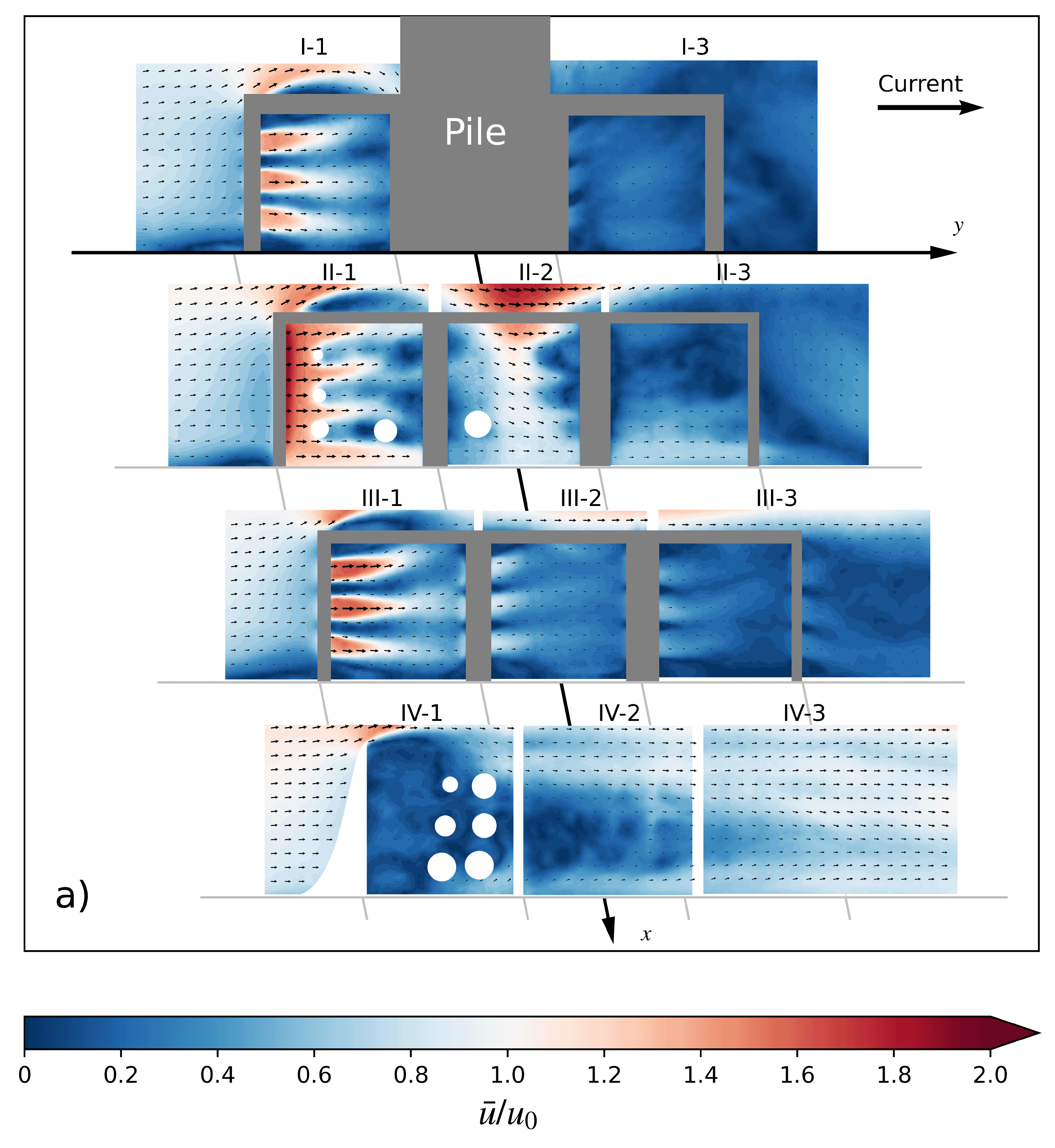}
\end{figure}
\begin{figure}
    \centering
    \includegraphics[width=\linewidth]{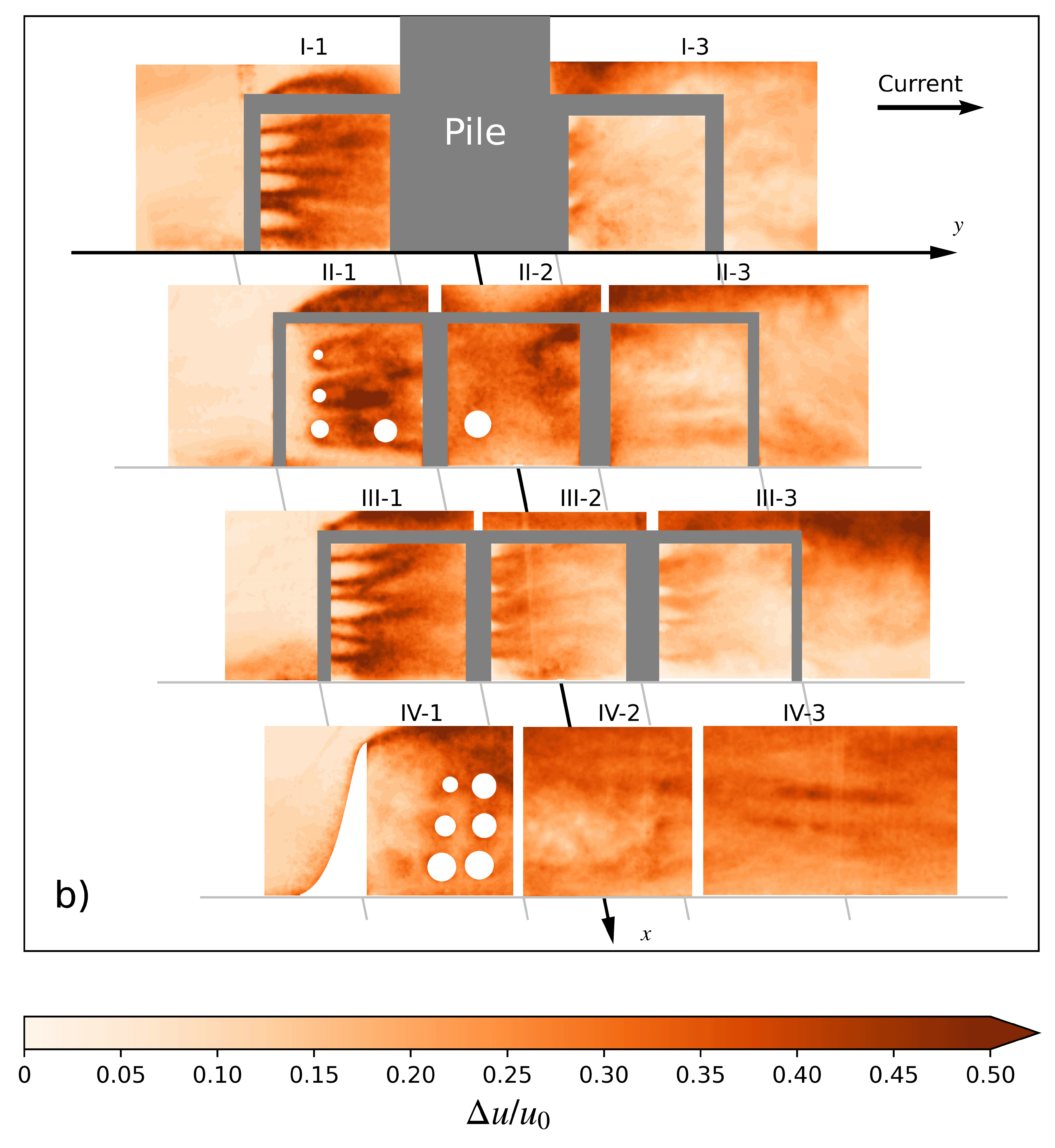}
  \caption{PIV measurement for test F2: (a) mean flow and (b) unsteadiness.}
  \label{fig_9_cubic}
\end{figure}

Panel I reveals the flow inside the C-AR placed upstream and downstream of the pile. In FOV I-1, the flow enters the C-AR through the 3$\times$3 holes, forming the three 'red fingers'. The velocity in these 'fingers' is approximately 1.3$u_0$, suggesting areas with higher flow speed. As the high-speed jets enter the holes, they are dissipated and mixed with the surrounding low-speed flow, so $\bar{u}$ becomes more spatially uniform and less than $u_0$ near the downstream part of C-AR. The unsteadiness $\Delta u$ in I-1 is mainly distributed between the 'fingers', which are regions with strong mean velocity gradients due to the spatial heterogeneity of $\bar{u}$. This shear motivates turbulence production, leading to much greater turbulent kinetic energy compared to that in FOV I-1 of control test in Fig.~\ref{fig_9_control}. Because turbulence represents flow’s capability of suspending sand, the results on $\Delta u$ indicate that some scour may occur beneath the upstream row of C-ARs. The $\bar{u}$ in FOV I-3, similar to that in Fig.~\ref{fig_9_control}, remains low, but the unsteadiness $\Delta u$ becomes much weaker. This is because the presence of the C-ARs prevents the formation of a flow separation zone behind the pile. However, strong $\Delta u$ is observed above the C-AR (the dark red zone in I-3), showing that a conventional turbulent wake behind the pile still exists at higher levels. In other words, the C-ARs indeed create a 'quiet zone' with both low $\bar{u}$ and $\Delta u$ behind the pile.

Panel II shows the flow in the gap between two streamwise rows of C-AR. In FOV II-1, the primary feature is the high $\bar{u}$ zone extending from the upstream edge, often referred to as the 'narrow gap' effect. Four distinct 'fingers' are formed in this zone, corresponding to the four completely blocked horizontal layers (i.e., layers without holes) on the lateral face of the C-ARs. Between these 'fingers', several white masks indicate locations of holes on the lateral face of C-ARs, where intense lateral flows are generated, suggesting that high-speed flow in the gap escapes through these holes, leading to a reduction of flow velocity. Notably, a 'finger' appears just above the sand bed, indicating that the accelerated flow within the gap can interact with the sand bed and potentially cause erosion. Similar to I-1, $\Delta u$ is concentrated in the regions between the high-speed 'fingers'. FOV II-2 depicts the flow around the side of the pile, where the $\bar{u}$ is greatly reduced due to the sheltering effect of C-ARs, compared to the control case. Because of the blockage of the front rows, most of the incoming water is diverted above or around the C-AR array, leaving only a small portion of the flow to reach the pile. As a result, scour around the lateral sides of the pile is expected to be reduced. In FOV II-3, the wake flow is also considerably weaker in terms of both $\bar{u}$ and $\Delta u$, which again demonstrates the ability of C-AR to create a quiet wake zone.

Panel III presents the flow through the outer streamwise row of C-ARs. The flow in FOV III-1 closely resembles that in I-1, showing high-velocity 'fingers'. The flow in both III-2 and III-3 has low $\bar{u}$ and $\Delta u$, suggesting that the flow is drastically attenuated after passing through the upstream rows of C-AR. 

Panel IV shows the flow immediately outside the C-AR array's lateral boundary. Incoming flows are deflected laterally around the ARs by the front face. The white circular masks cover the holes on the lateral face of the front AR. Flow inside the AR escapes through these holes, creating high out-of-plane velocity. FOVs IV-2 and IV-3 clearly show that the main flow bypasses the C-AR array and gradually reorganizes itself downstream, i.e., the high $\bar{u}$ zone is gradually extended downward and eventually occupies the entire depth of FOV near the downstream end of IV-3.

\subsection{Flows with H-ARs}
\begin{figure}
  \centering
    \includegraphics[width=\linewidth]{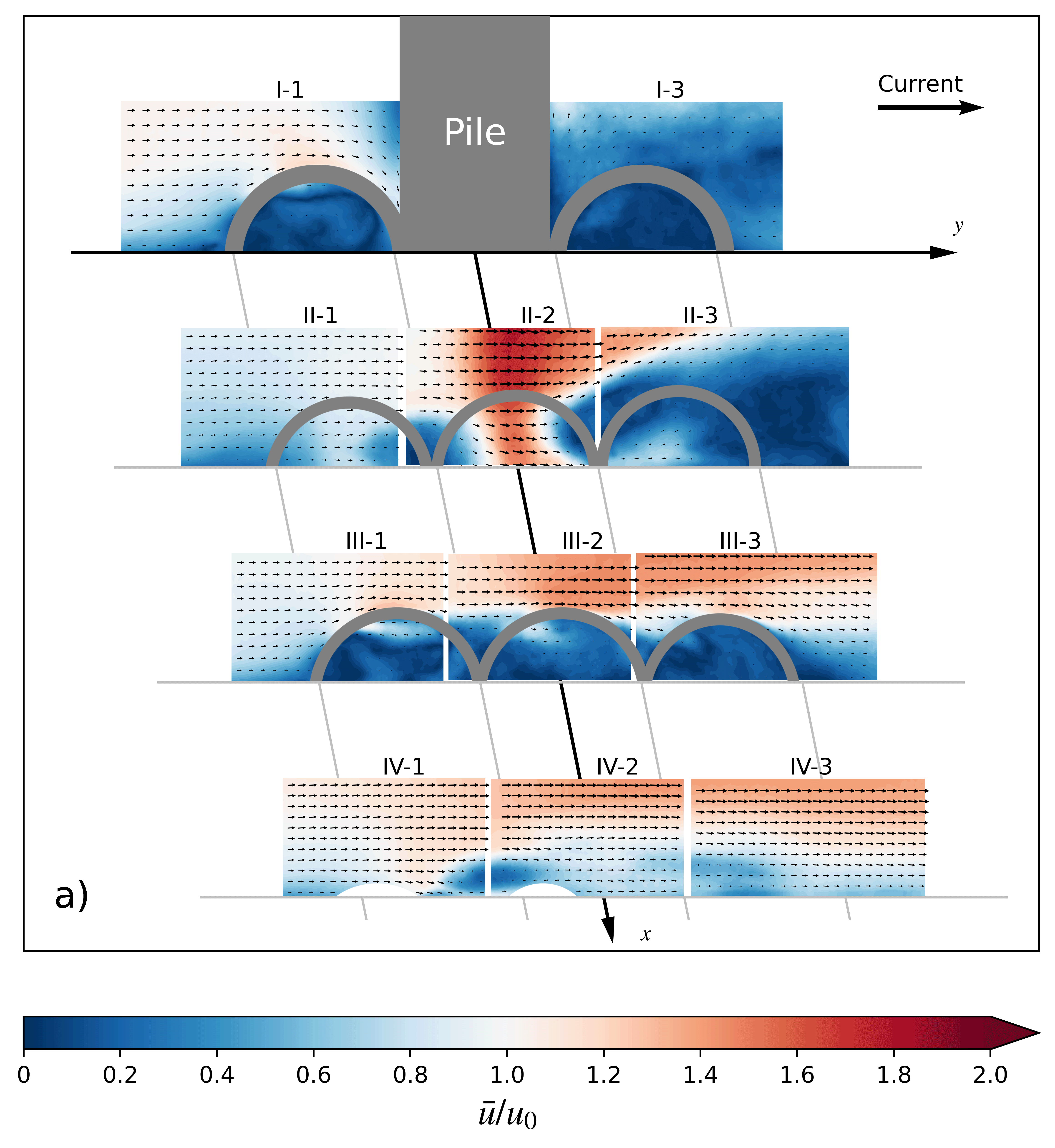}
\end{figure}
\begin{figure}
    \centering
    \includegraphics[width=\linewidth]{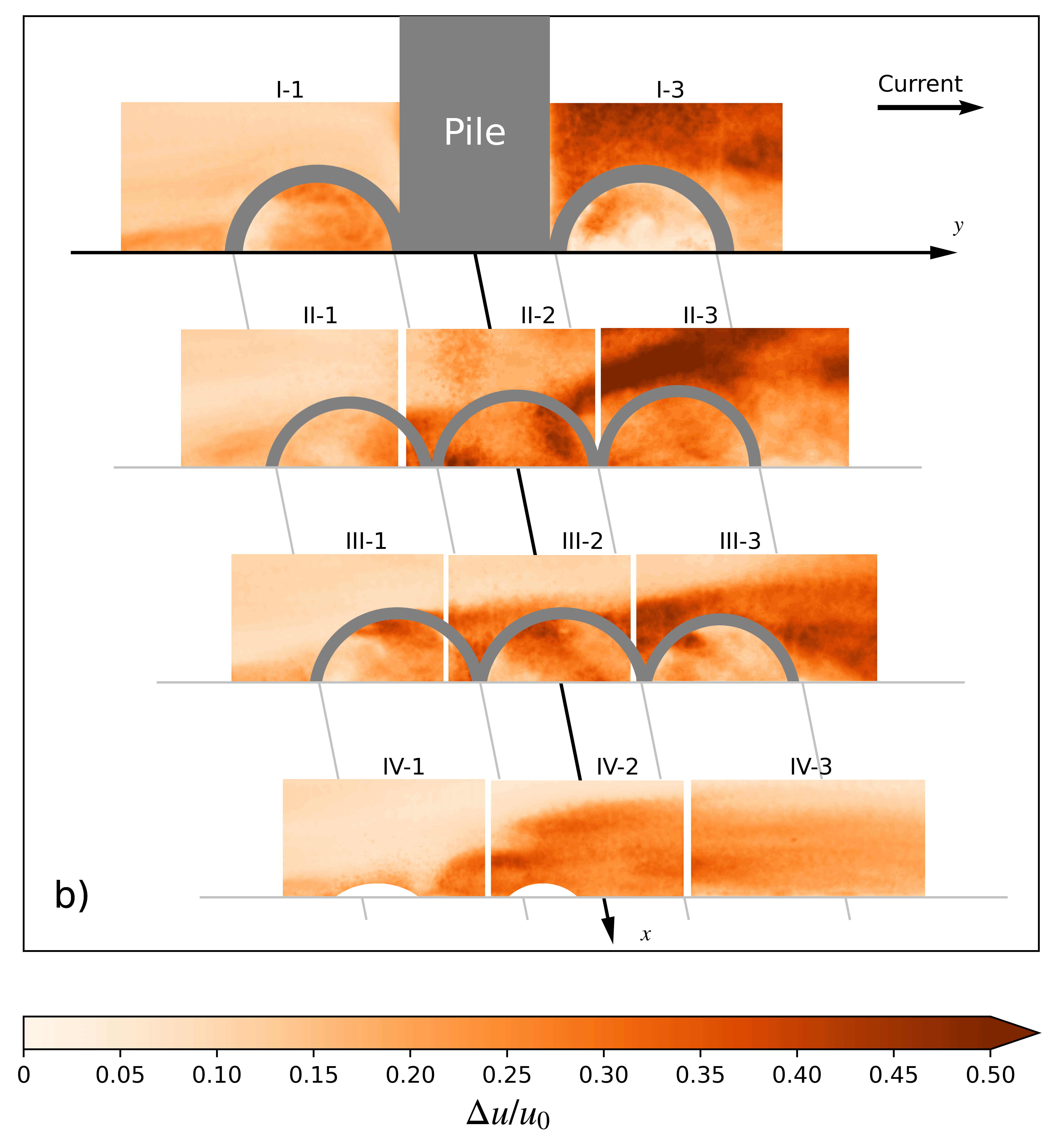}
  \caption{PIV measurement for test F3: (a) mean flow and (b) unsteadiness.}
  \label{fig_9_hemisphere}
\end{figure}

The flow fields with H-ARs share some similarities with those of C-ARs. To begin with, the flow in the wake region is drastically reduced for both $\bar{u}$ and $\Delta u$, except for some minor differences due to the shape of ARs. In addition, the flows entering the ARs through the holes always have strong velocities and triggers internal turbulence. Below, we summarize some unique features of H-ARs.

First, an enhanced downward flow develops along the upstream side of the pile, as shown in FOV I-1 of Fig.~\ref{fig_9_hemisphere}. The incoming flow is guided by the curved surface of the upstream H-AR unit, amplifying the downward flow in front of the pile, which may lead to scour on the upstream side.

Second, the flow around the lateral side of the pile is further amplified. A comparison of FOV II-2 across all three cases shows that the H-AR configuration gives the highest $\bar{u}$ at II-2, which is up to 1.6$u_0$. This flow can contribute to scour on the lateral side of the pile. In addition, there is a high $\Delta u$ zone right above the sand bed in II-2, suggesting the presence of a horseshoe vortex. Nevertheless, the near-bed $\Delta u$ in II-2 remains weaker than that in the control test, so some reduction of lateral-side scour is still expected.

Third, the difference in FOV II-1 between C-ARs and H-ARs reveals that the 'narrow gap' effect prominent in the C-AR case is absent in the H-AR case. This is attributed to the curved surface of H-AR, which limits the 'narrow gap' effect. Consequently, we can anticipate less scour within the gaps among H-ARs, especially between the front-row ARs.

\subsection{Brief summary and discussion}
In summary, the flow field gained from the control test is well consistent with those phenomena presented in Fig.~\ref{fig_2}, confirming the accuracy of PIV measurements. For C-ARs, high-speed flow and turbulence are primarily concentrated within the front-row units, indicating a high potential for local scour. Strong velocity is observed within the gaps between C-AR units, suggesting the narrow-gap effect. H-ARs have less blockage of incoming flow. A strong downward flow in front of the pile is formed with the guidance of H-AR's curved surface. In both tests with ARs, flow velocity and unsteadiness are reduced by 50$\sim$80\% in the downstream wake zone, meaning the areas behind the pile are likely to be effectively protected from scour. These hypotheses will be verified in the live-bed tests shown in the next section.

\section{Live-bed tests}

This section presents the experimental results of live-bed tests, which demonstrate the feasibility of using ARs to reduce monopile scour under currents. Some key observations are first summarized to give the reader an overall picture. AR's efficiency of scour protection, quantified by the reduction of scour depth and width, is presented subsequently. Finally, modifications of AR array layout were explored.

\subsection{Key observations}

\begin{figure}
  \centering
  \includegraphics[width=\linewidth]{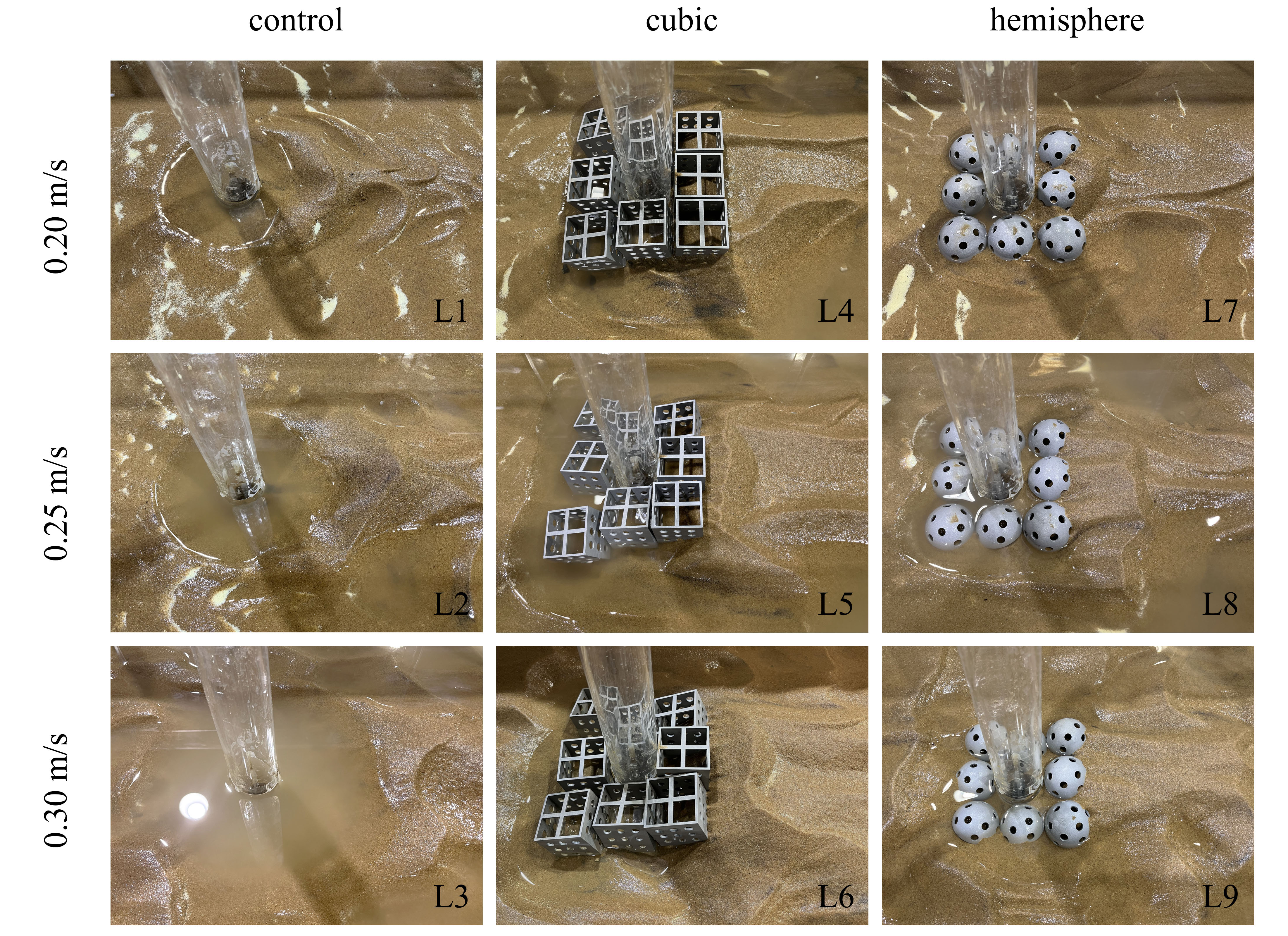}
  \caption{Photos of the scout pit at the end of all live-bed tests. Each row has the same current velocity, and each column has the same scour protection (or no protection).}
  \label{fig_11}
\end{figure}

The photos of the scour pit at the end of each live-bed test are presented in Fig.~\ref{fig_11}. In the three control tests (L1-L3), a typical current-induced scour around a monopile is produced. The shape of the scour pit is circular and is centered at the monopile. The radius of the circle apparently increases with the flow condition, $u_c$, i.e., from about $1.5D$ under $u_c$ = 0.20 m/s to about $2.5D$ under $u_c$ = 0.30 m/s. Deposition of sand occurs behind the circular-shaped scour, and the size of this deposition zone also increases with the size of the scour pit. 

In the three C-AR tests, the main feature is that some C-ARs were moved a bit, so the closely packed array (see Fig.~\ref{fig_5}) became loose. Some scour around the edge of the AR array occurred, which could exceed the scour depth at the monopile. There was some topological change underneath the ARs, so the C-ARs eventually sat on an uneven surface, which led to some tilting or displacement of the C-AR units. The maximum displacement of C-AR was up to half the reef's width. The front row of C-ARs experienced the most significant topological change locally, and the two corner units of the front row had the largest displacement. It will be shown later that the breakdown of the front row will reduce the efficiency of scour protection. In the downstream of the AR array, the deposition zone becomes wider and contains more sand than that in the control tests.

In the three H-AR tests, the AR units remained closely packed as in the initial setup, regardless of the flow condition. Compared to the C-AR tests, there was less edge scour around the AR units, so topological change underneath the ARs was not as dramatic as in the C-AR tests. Increasing the current speed actually gives more scour around the monopile, so the H-AR units appeared to 'sink' into the sand bed and fill the scour pit. There was also a significant deposition zone behind the array.

\begin{figure}
  \centering
  \includegraphics[width=\linewidth]{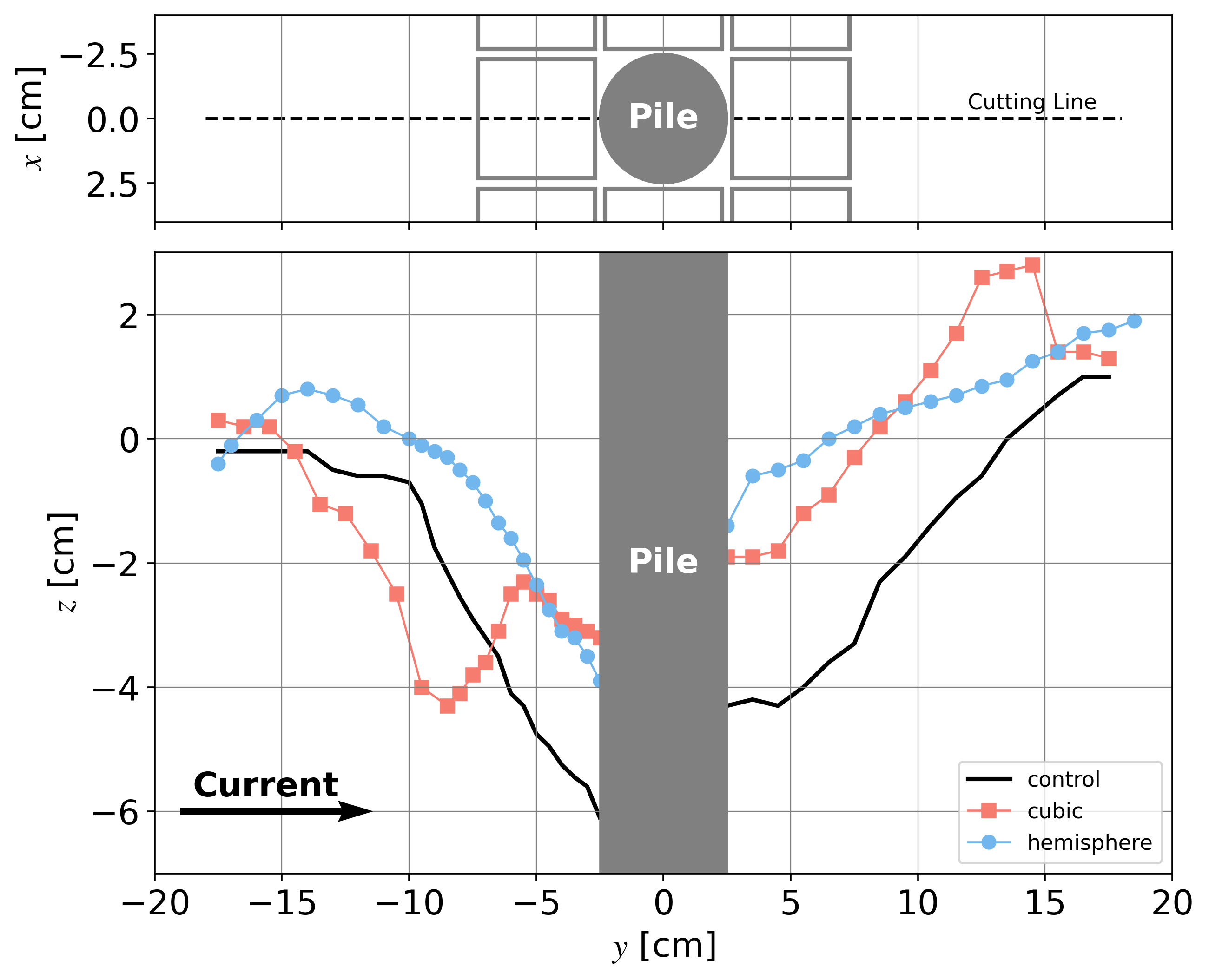}
  \caption{Profiles of scour pit along the streamwise centerline cut for tests L3, L6, L9 (current velocity is 0.30 m/s).}
  \label{fig_13}
\end{figure}

To quantitatively show the influence of ARs on the scour process, change of bottom elevation, $\Delta z$, was measured along the streamwise centerline cut for tests L3, L6, and L9, which have a current velocity of 0.30 m/s, and the results are presented in Fig.~\ref{fig_13}.
In the control test (black curve), the scour depth increases from the far fields toward the pile, and the largest scour depth occurs at the upstream side of the pile ($\Delta z$ = $-$6.1 cm). At about 13 cm or 2.5$D$ downstream, $\Delta z$ becomes positive, which marks the beginning of the deposition zone. 
After applying the ARs, there was a notable reduction of scour depth in the vicinity of the pile, suggesting that ARs indeed reduce scour.

On the upstream side, C-ARs gave $\Delta z$ = $-$3.2 cm at the monopile, which is better than the H-ARs ($\Delta z$ = $-$3.9 cm). As found in the fixed-bed tests, H-ARs guide a stronger descending flow in front of the pile, which is absent when C-AR is present. The region of $y$ = $-$7.5 cm to $y$ = $-$2.5 cm is where the front units of AR sit. In this region, $\Delta z$ of the H-AR test decreases towards the pile, which forms a slope that tilts the H-ARs inward (or toward the pile, see the photo in Fig.~\ref{fig_11}). $\Delta z$ of the C-AR test has a local maximum ($\Delta z$ = $-$2.3 cm) at about $y$ = $-$5.5 cm, which is within the C-AR. This means the C-AR traps some sand within it. The minimum $\Delta z$ of the C-AR test occurs at $y$ = $-$8.5 cm, which is within the edge scour outside the upstream front of C-AR array. It is interesting to see that this edge scour is much larger than the scour at the monopile. This agrees with the findings from fixed-bed tests, i.e., high-speed flow and intense turbulence are concentrated in the front-row area of the C-AR array. In other words, the C-ARs 'protect' the pile by 'sacrificing' the first-row unit. The large edge scour also makes the first-row unit tilting outward from the pile (see the photo in Fig.~\ref{fig_11}). The upstream width of the scour (i.e., the $y$ distance of the point with $\Delta z$ = 0 to the pile) is reduced in the H-AR test (i.e., $\Delta z$ = 0 at $y$ = $-$10 cm) but remains almost unchanged in the C-AR test (i.e., $\Delta z$ = 0 cm at $y$ = $-$15 cm).  

The reduction of scour is much more significant on the downstream side. Right at the pile's edge ($y$ = 2.5 cm), both ARs can reduce the scour from $\Delta z$ = $-$4.3 cm to $\Delta z$ = $-$1.4 cm or $-$1.9 cm, i.e., a roughly 60\% reduction. The scour width is also drastically reduced. $\Delta z$ = 0 occurs at $y$ = 13.5 cm in the unprotected test, but is reduced to about $y$ = 7 cm in both AR tests. The C-ARs also give more deposition in the wake. The pattern of scour reduction agrees with the findings in fixed-bed tests, i.e., both mean flow speed and the time fluctuations are greatly damped by the ARs in the wake region. The C-AR is twice taller than the H-AR, so it can better reduce the flow and allow more deposition of sand in the wake.

\subsection{Efficiency of scour protection}

\begin{figure}
  \centering
  \includegraphics[width=0.7\linewidth]{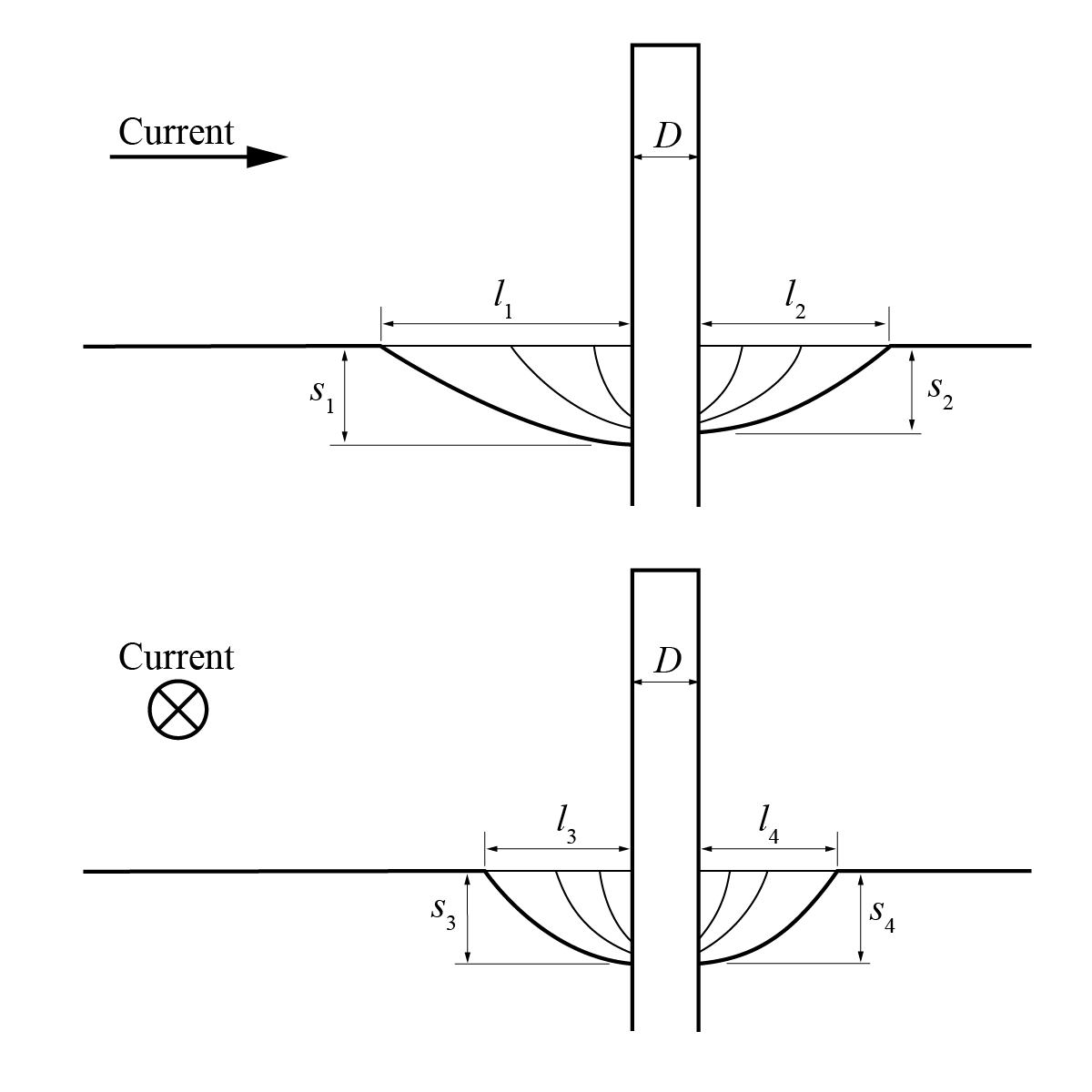}
  \caption{Definition of scour depth and width.}
  \label{fig_7}
\end{figure}

To quantify the AR's efficiency of scour protection, we introduced some geometric parameters, i.e., the scour depth and width at the upstream, downstream, and lateral sides of the pile, which are defined in Fig.~\ref{fig_7}. 
Scour depths around the pile were acquired by reading some rulers attached vertically to the model pile. The scour widths were measured using a long ruler. The edge of the scour pit can be visually identified (see the photos in Fig.~\ref{fig_11}). The measurement error of scour width is mainly the error of identifying the scour edge, which is estimated to be less than 0.5 cm or about 5\% relative error. The normalized scour depth $ s/s_0 $ and width $ l/l_0 $ are introduced as the final experimental results, where ($s_0$, $l_0$) represent the scour depth and width of the control tests and ($s$, $l$) are the scour depth and width with ARs. Since the two lateral sides are symmetric, the averaged results are retained. Detailed data is shown in Table.~\ref{tab_scour}.

\begin{table}
  \centering
  \caption{Measurements of scour depth and width around the pile.}
  \begin{tabular}{ccccccc}
    \hline
    Test & $s_1$ [cm] & $s_2$ [cm] & $mean(s_3,s_4)$ [cm] &  $l_1$ [cm] & $l_2$ [cm] & $mean(l_3,l_4)$ [cm] \\
    \hline
    L1 & 3.0 & 1.5 & 3.0 & 5.5 & 5.0 & 4.2 \\
    L2 & 5.1 & 3.2 & 4.6 & 7.2 & 9.0 & 7.1 \\
    L3 & 6.1 & 4.3 & 5.7 & 10.5 & 10.8 & 8.5 \\
    \hline
    L4 & 0.0 & 0.0 & 1.8 & 7.7 & 0.0 & 12.0 \\ 
    L5 & 3.2 & 0.0 & 3.2 & 10.5 & 0.0 & 15.0 \\
    L6 & 3.2 & 1.9 & 4.1 & 12.2 & 5.0 & 15.3 \\
    L7 & 1.9 & 0.0 & 2.3 & 3.5 & 0.0 & 4.1 \\
    L8 & 2.9 & 0.0 & 2.9 & 6.0 & 0.0 & 8.0 \\
    L9 & 3.9 & 1.4 & 2.7 & 8.9 & 4.1 & 7.5 \\
    \hline
    L10 & 3.5 & 1.7 & 3.7 & - & - & - \\
    L11 & 4.2 & 1.6 & 4.3 & - & - & - \\
    L12 & 2.2 & 0 & 1.6 & - & - & - \\
    L13 & 2.0 & 0.9 & 2.4 & - & - & - \\
    \hline
  \end{tabular}
  \label{tab_scour}
\end{table}

\begin{figure}
  \centering
  \subfloat[Test L1, L4, L7]{
    \includegraphics[width=0.48\textwidth]{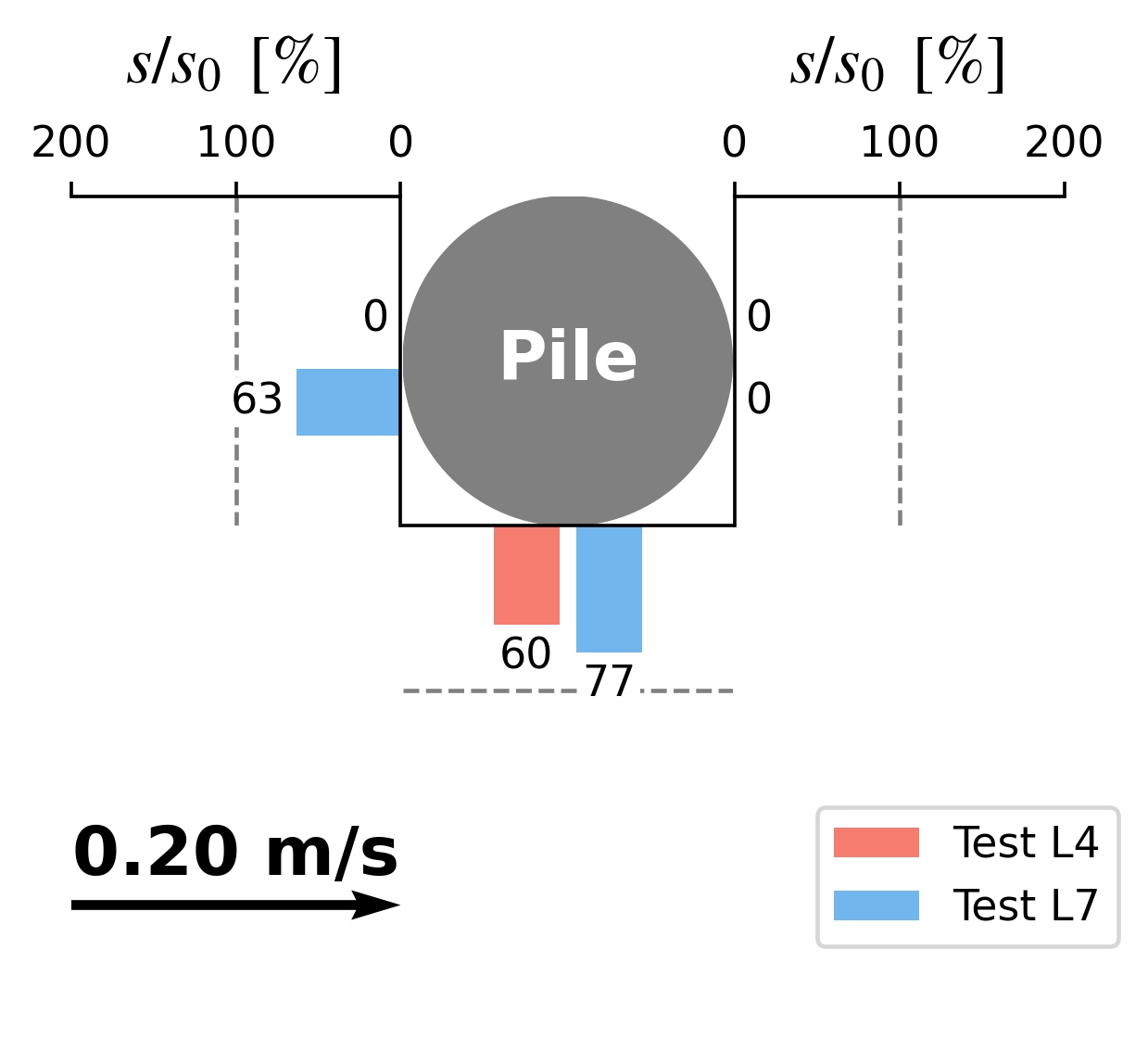}
  }
  \hfill
  \subfloat[Test L2, L5, L8]{
    \includegraphics[width=0.48\textwidth]{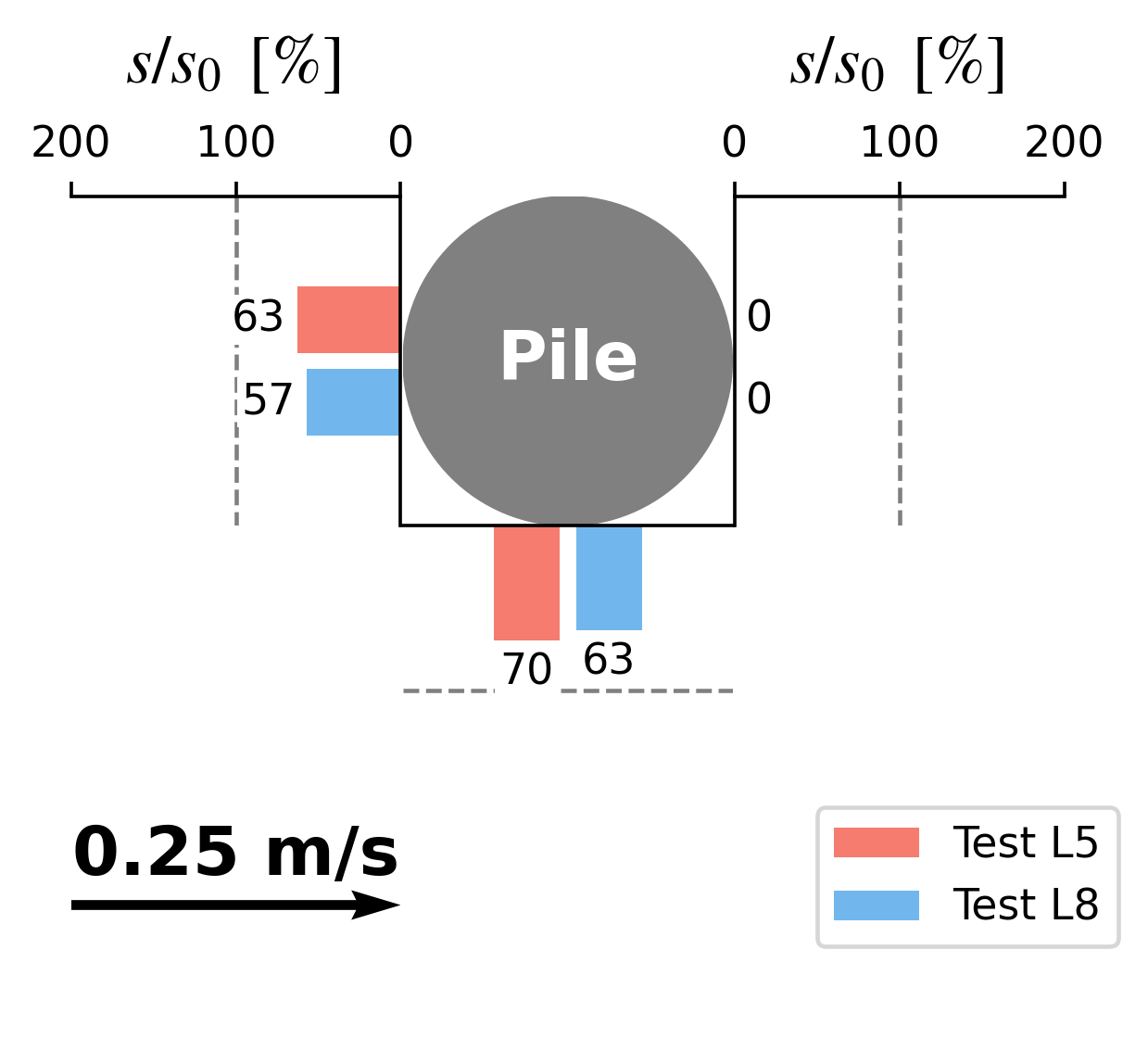}%
  }
  
  \subfloat[Test L3, L6, L9]{
    \includegraphics[width=0.48\textwidth]{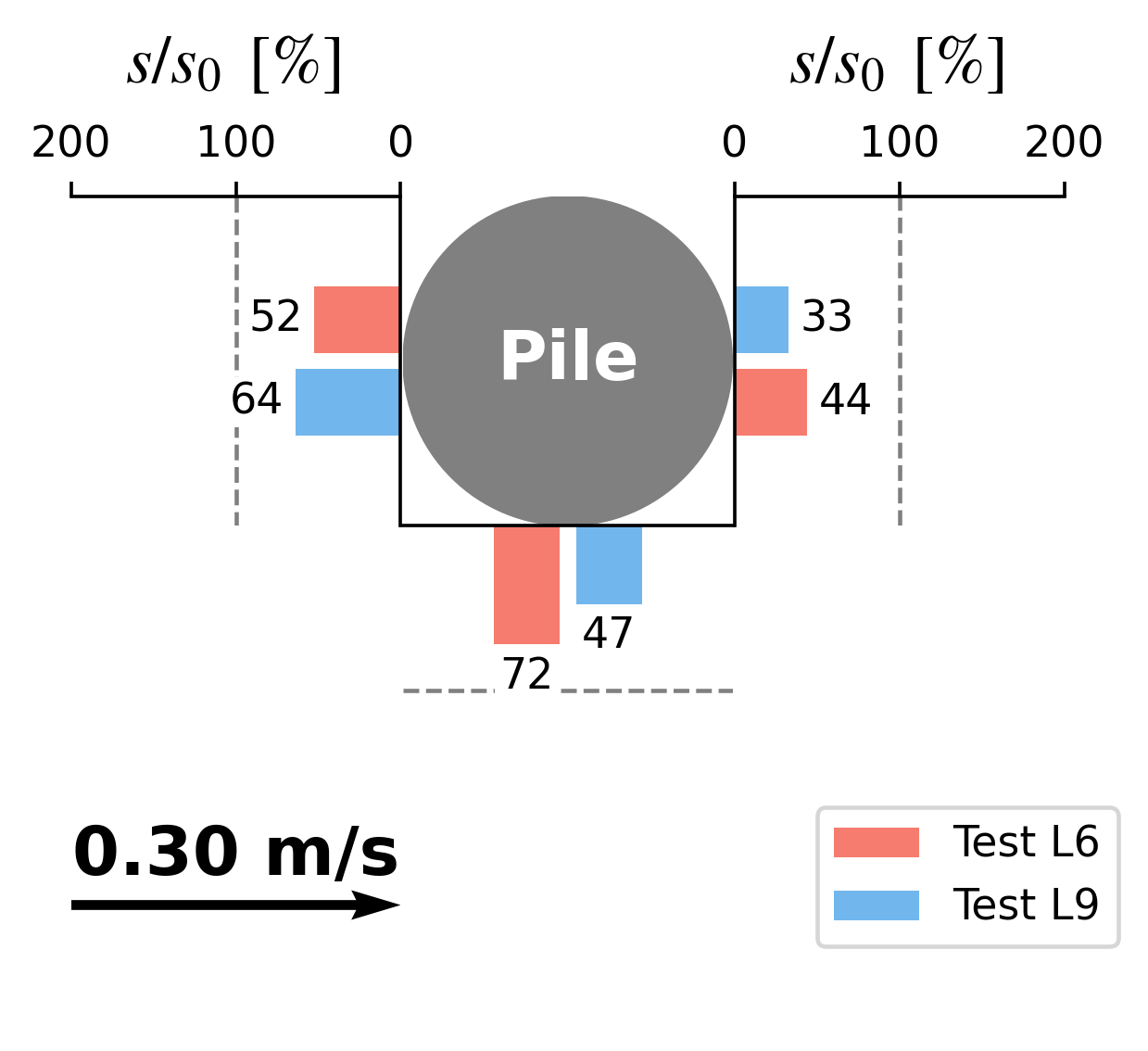}%
  }
  \caption{Scour depth of the live-bed tests L1 to L9.}
  \label{fig_12_d}
\end{figure}

The dimensionless scour depths at the pile are compared in Fig.~\ref{fig_12_d}. For the low-flow group (L1, L4, and L7, $u_c=$ 0.20 m/s), the AR array efficiently reduced the scour. With the C-AR array (test L4), there was no scour ($s/s_0=$ 0) at the upstream and downstream sides, and only the lateral side had 60\% scour. As shown in Fig.~\ref{fig_11}, the C-ARs in test L4 remain closely packed around the pile, due to relatively less edge scour, so they can efficiently reduce the flow around the pile, which have been illustrated by the fixed-bed test. Meanwhile, as shown in Fig.~\ref{fig_9_cubic} (FOVs II-1 and II-2), the gap between two streamwise rows of C-ARs forms a narrow gap for the flow to go through, so there is a strong near-bottom flow sweeping the lateral side of the pile. In test L7 with the H-ARs, the upstream and side scour is reduced only to 63\% and 77\% of those in the control tests. This relatively poor performance can be explained by the sizable flows in these areas revealed in the fixed-bed tests. Nevertheless, the H-ARs can still prevent scour in the wake region under such low flow condition. 
By increasing the flow condition, the efficiency of C-ARs was reduced. The upstream scour depth was increased to 50\% to 60\% of the original value, while 33\% scour in the downstream occurred under the strongest current. The efficiency of H-ARs remained mostly unchanged, except that 44\% scour in the downstream occurred under the strongest current. The difference can be attributed to the noticeable displacement and tilting of C-ARs discussed in the last subsection, which opened the gaps between C-AR units for flow to go through and hence reduced the ARs' protection. The closely packed layout of H-ARs, however, was almost not changed, so they can provide similar protection under varied flow conditions.  

\begin{figure}
  \centering
  \subfloat[Test L1, L4, L7]{
    \includegraphics[width=0.48\textwidth]{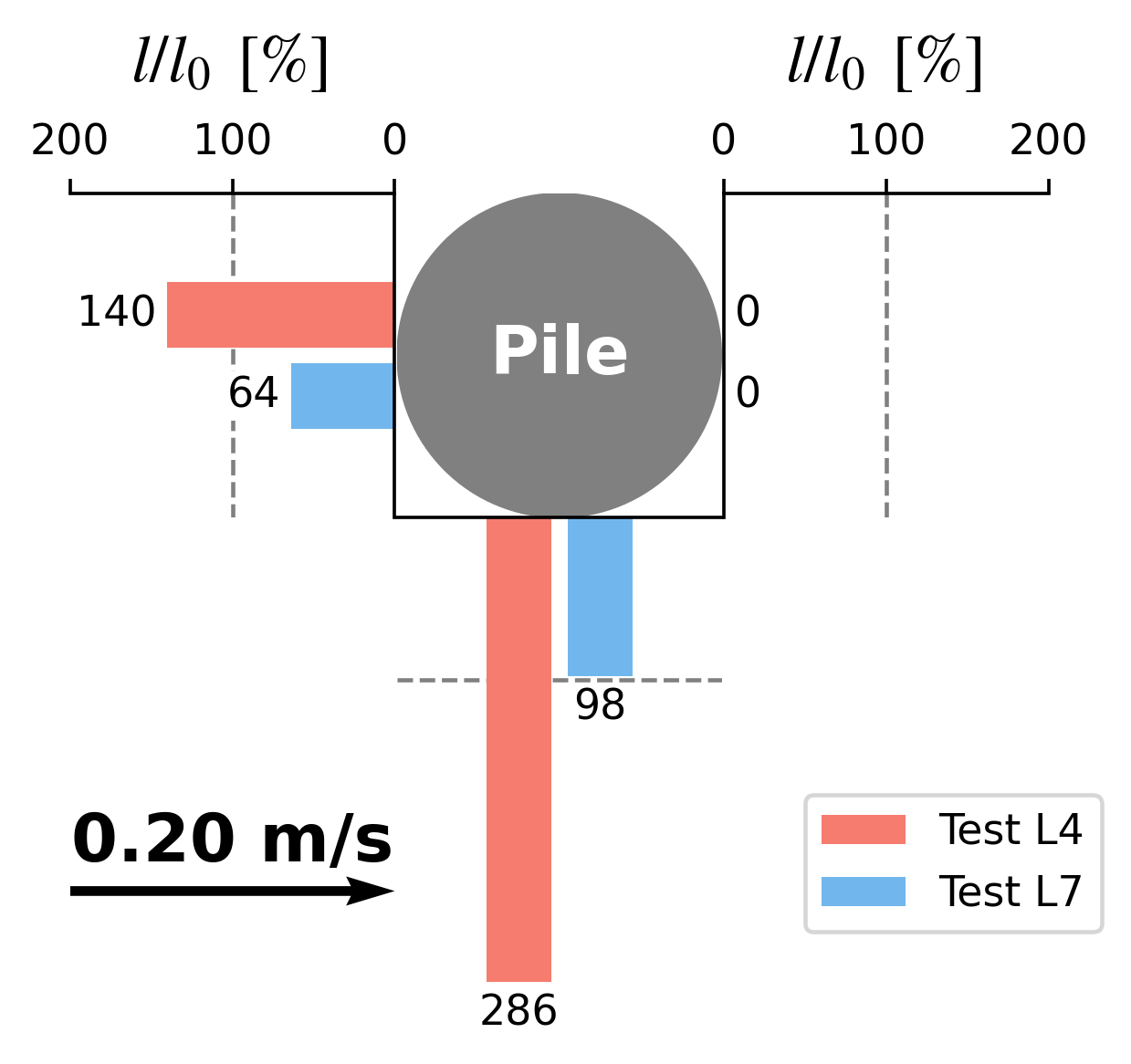}
  }
  \hfill
  \subfloat[Test L2, L5, L8]{
    \includegraphics[width=0.48\textwidth]{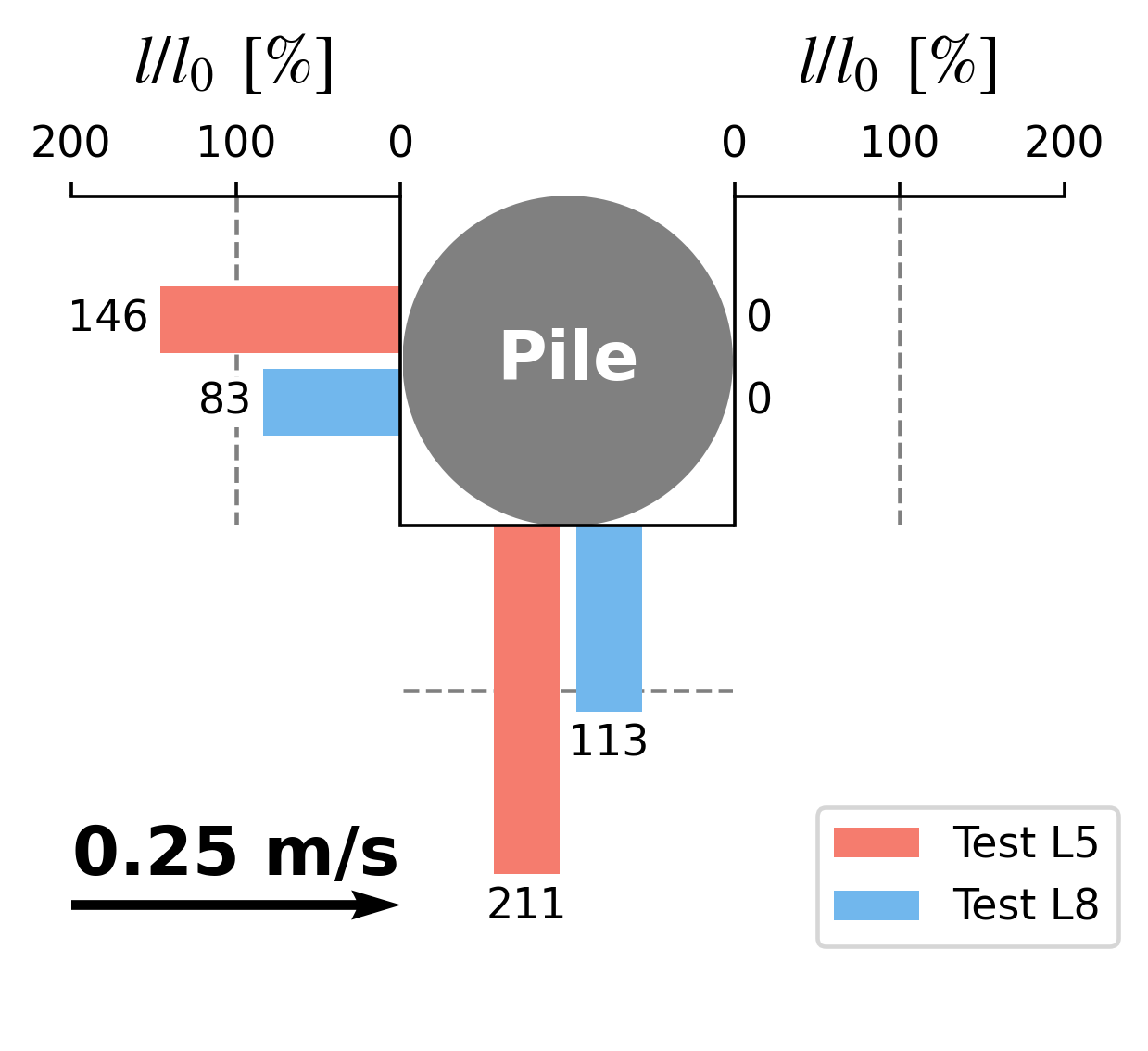}%
  }
  
  \subfloat[Test L3, L6, L9]{
    \includegraphics[width=0.48\textwidth]{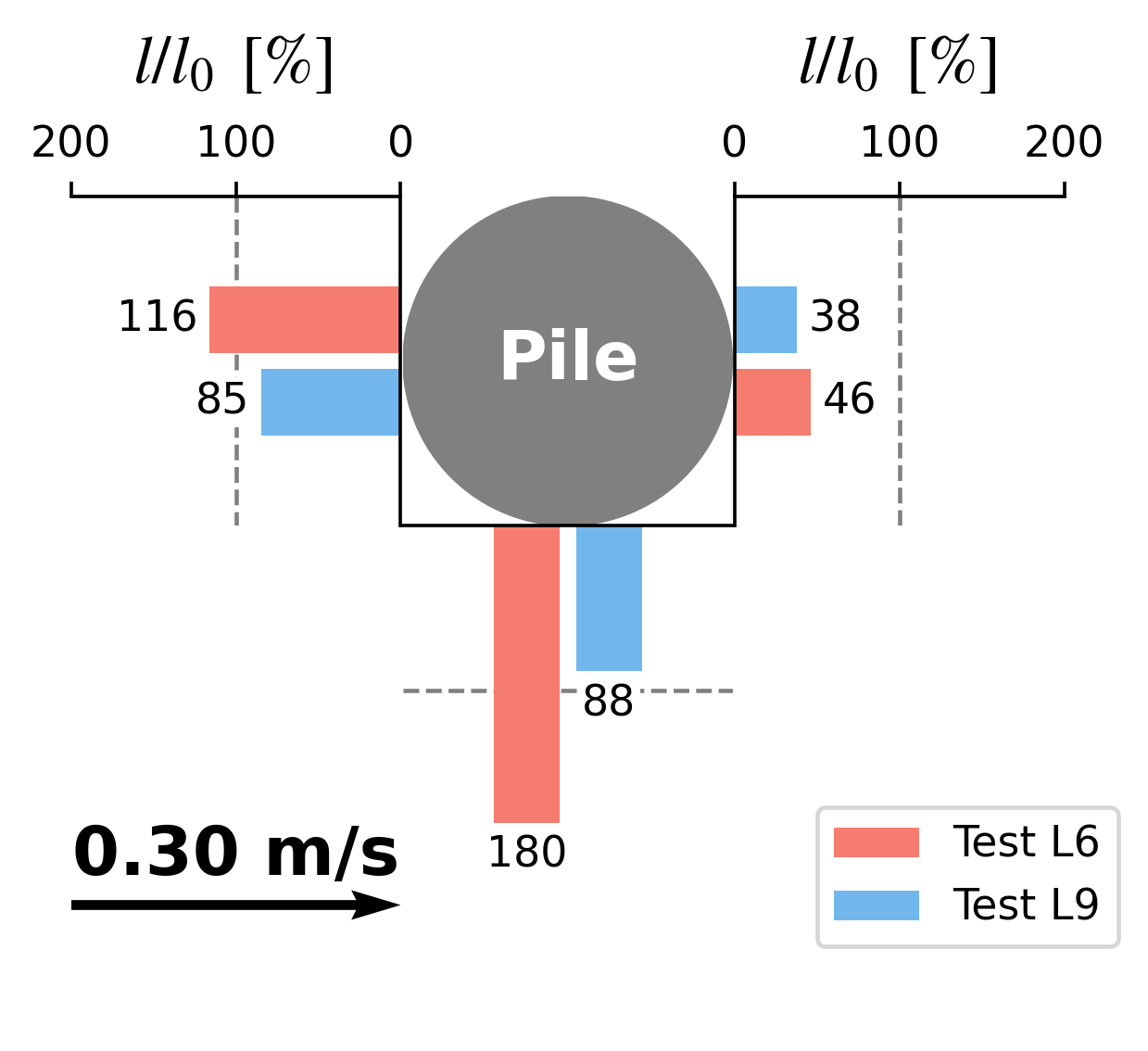}%
  }
  \caption{Scour width of the live-bed tests L1 to L9.}
  \label{fig_12_w}
\end{figure}

The dimensionless scour width are shown in Fig.~\ref{fig_12_w}. It is evident that C-ARs significantly widened the scour pit at both the front and lateral sides, but H-ARs can reduce the scour width slightly (about 10\% to 20\%). C-ARs themselves have strong edge scours, so they naturally expand the region with scour. It should be noted that the increase of lateral scour width is up to twice the increase of upstream scour width. The blockage of the C-AR array diverts the incoming flow to the two lateral sides of the whole array, which can be partly demonstrated by the white masks (i.e., strong out-of-plane flow velocity) in panel IV of Fig.~\ref{fig_9_cubic}a). As a result, the lateral-side seabed was heavily eroded, leading to the significant lateral edge scour. The reduction of scour width with H-ARs can be explained as follows. Since these units do not have strong outside edge scour, the overall shape of the scour pit remains as an inverse cone. As the scour pit develops, the H-AR units essentially fill the pit, which hinders the increase of both scour depth and width. 

In summary, the flow characteristics observed from fixed-bed tests can account for most phenomena in the live-bed tests. Seabeds exposed to higher flow velocity tend to suffer more severe scour. On the contrary, areas with low flow velocity typically showed minimal scour and may even develop deposition zones. In addition, ARs' tilting and displacement, which are closely related to the intensity of edge scour, also play a crucial role in scour protection. If the AR array is broken apart, the unprotected seabeds in the resulting gap are likely to be eroded.

\subsection{Modification of AR layout}

The simple layout of AR in the previous subsections can be further modified. The three most possible changes are: (1) increasing the spacing among the AR units, (2) adding more rows of ARs, and (3) integrating different AR types. Five additional live-bed tests, L10$sim$L14 in Table~\ref{tab_live-conditions}, were conducted to investigate whether these changes can lead to positive results. Since the strongest flow condition gives the worst scour, these five tests were all conducted with a 0.30 m/s current. 

\begin{figure}
  \centering
  \includegraphics[width=\linewidth]{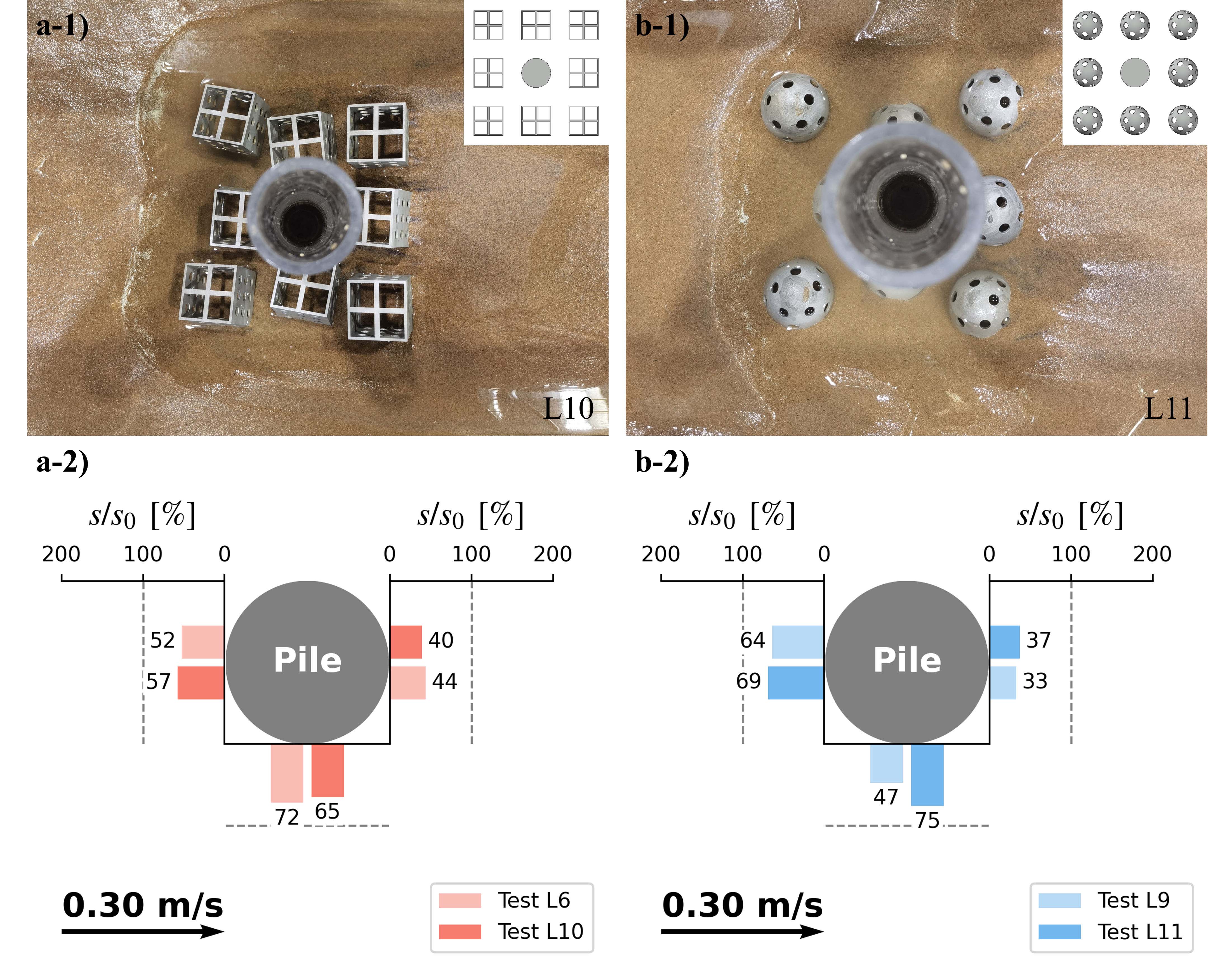}
  \caption{Experimental results with modified design of AR layout: increased spacing among ARs.}
  \label{fig_14}
\end{figure}

In tests L10 and L11, the spacing between two ARs is increased to 2.5 cm, or half of their dimension (diameter or edge length). The photos taken at the end of the test and the measured normalized scour depth are shown in Fig.~\ref{fig_14}. Increasing the spacing may reduce the 'narrow-gap' effect between ARs, thereby reducing the formation of high-speed jets that accelerate the scour.
However, the results are not positive. For the C-ARs, there was little change in scour depth compared to test L6 (Fig.~\ref{fig_14}a-2), which was expected, as the ARs in L6 were also separated in the end. For H-ARs, a greater scour depth was observed on all sides of the monopile (Fig.~\ref{fig_14}b-2), signifying a worse performance in scour protection. The increased gaps allow more flow to pass through, leading to greater scour, which outweighs the benefits of reducing the high-speed jets typically associated with narrow gaps.

\begin{figure}
  \centering
  \includegraphics[width=\linewidth]{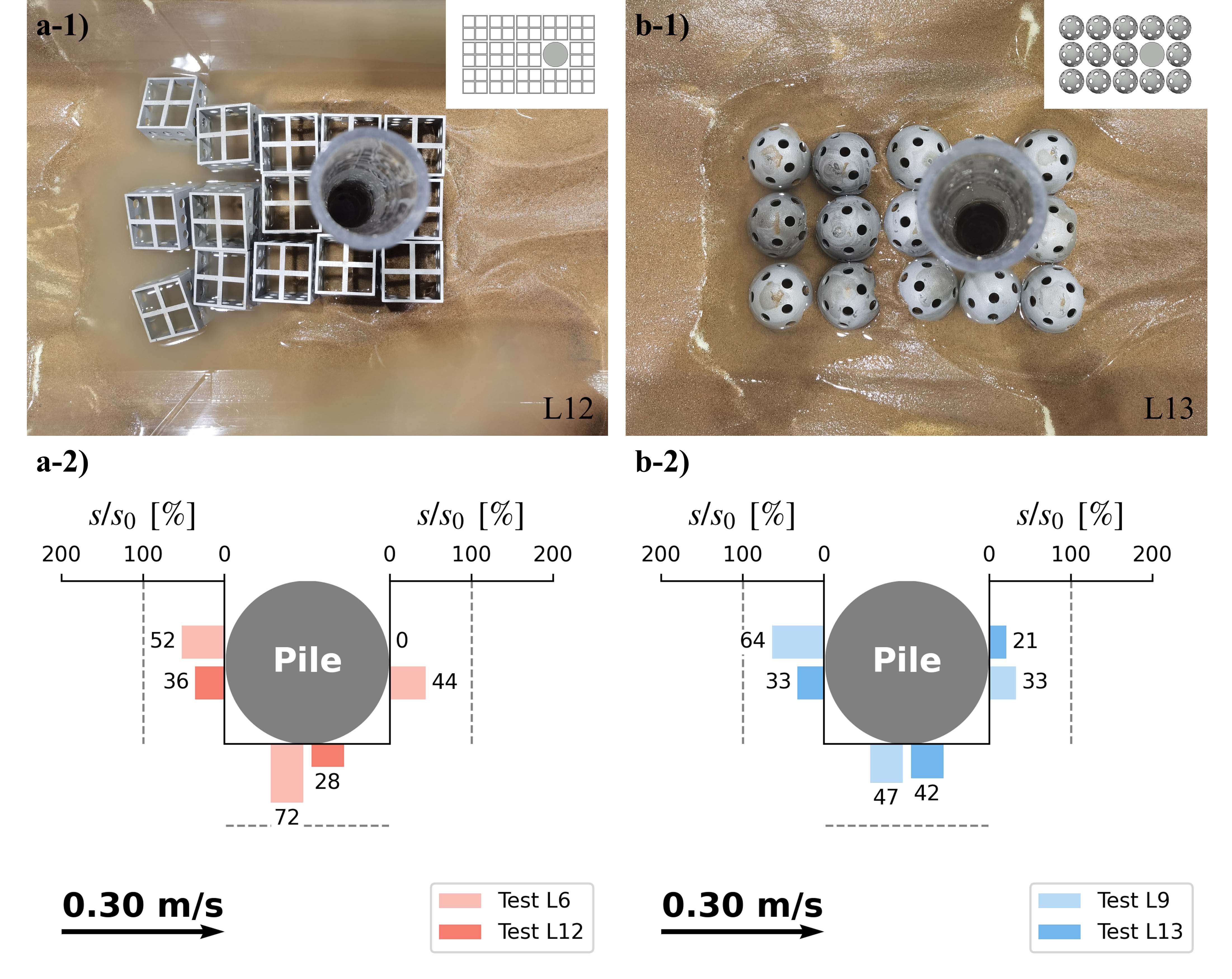}
  \caption{Experimental results with modified design of AR layout: adding two more rows of ARs in the upstream.}
  \label{fig_15}
\end{figure}

The second modified design involves adding another two rows of ARs in the upstream, as shown in Fig.~\ref{fig_15}. This design aims to further reduce the flow velocity that can reach the pile. Moreover, the area of typical scour protection generally covers the area of the scour pit that would arise \citep{RN35}, so a total of three rows were selected to completely cover the area of the scour pit observed in the control test. 
The results confirmed the effectiveness of this design. For C-ARs, although the newly added ARs were still displaced because of the strong edge scour, the ARs immediately around the pile can maintain their original positions (Fig.~\ref{fig_15}a-1). The scour depths around the pile were decreased (Fig.~\ref{fig_15}a-2): upstream scour from 52\% to 36\%, lateral scour from 72\% to 28\%, wake scour from 44\% to 0\% (i.e., an average total reduction of 79\%). This improvement demonstrates that C-ARs can provide excellent scour protection as long as they are tightly placed. The scour protection efficiency of H-ARs is also improved significantly (Fig.~\ref{fig_15}b-2) but still lower than C-ARs in general. In conclusion, adding multiple layers of ARs upstream is an effective strategy for enhancing scour protection performance.

\begin{figure}
  \centering
  \includegraphics[width=0.5\linewidth]{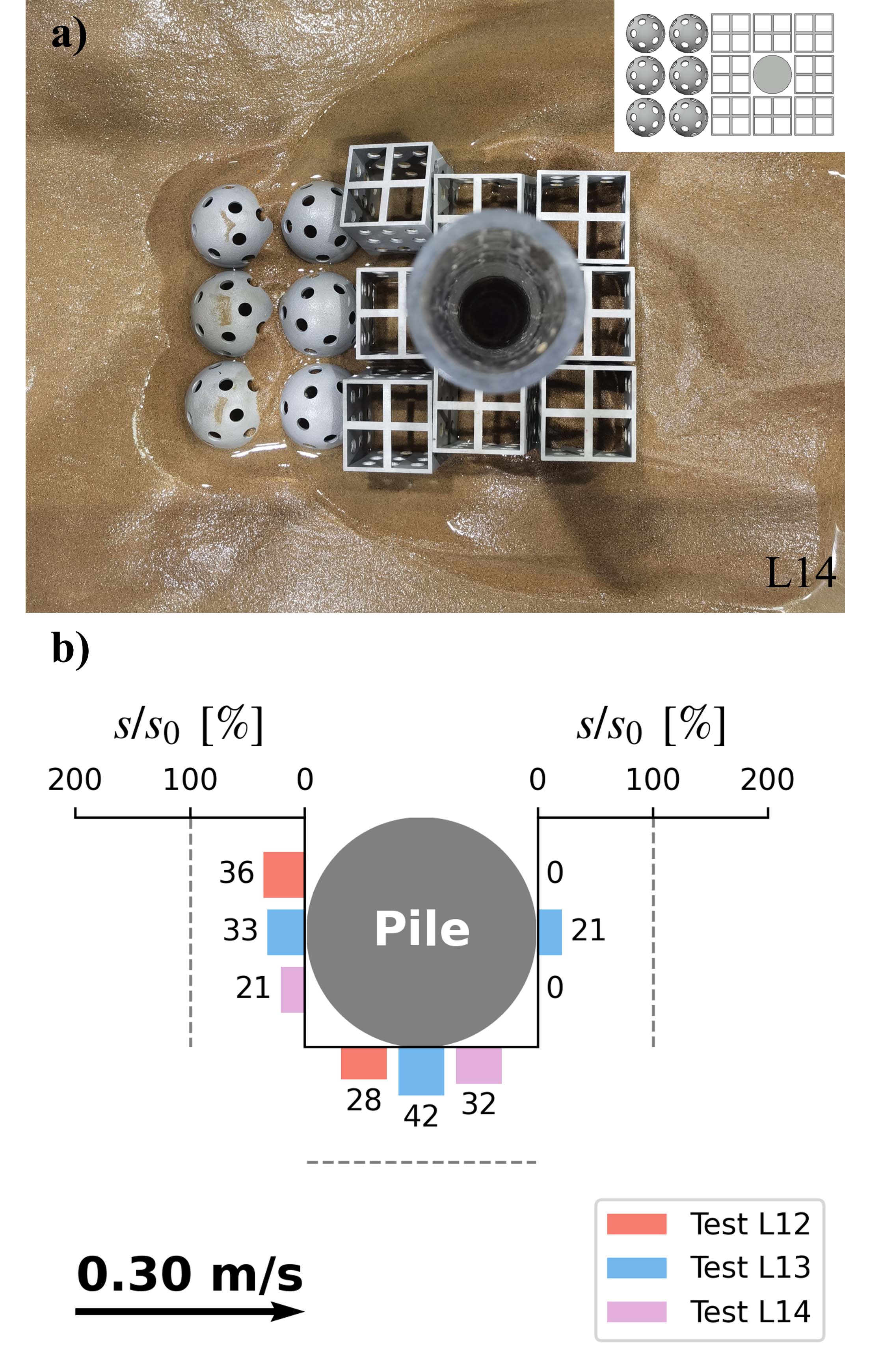}
  \caption{Experimental results with modified design of AR layout: combining the two AR types.}
  \label{fig_17}
\end{figure}

The third modification has two rows of H-ARs placed upstream from the 3$\times$3 C-AR array. This design combines the advantages of both ARs, i.e., H-AR has much less edge scour and C-ARs can better reduce scour near the pile if they remain intact.
The outcomes demonstrate a good success. As shown in Fig.~\ref{fig_17}a, the two corner C-ARs only tilt laterally a bit, indicating that the H-ARs indeed mitigate edge scour in front of the C-ARs. Consequently, the entire layout of ARs remains intact. The data presented in Fig.~\ref{fig_17}b validate the protective benefits of integrating the two AR types, i.e., scour depth reduction around the pile ranges from 70\% to 100\%.

\subsection{A suggested design of AR layout}

\begin{figure}
  \centering
  \includegraphics[width=0.7\linewidth]{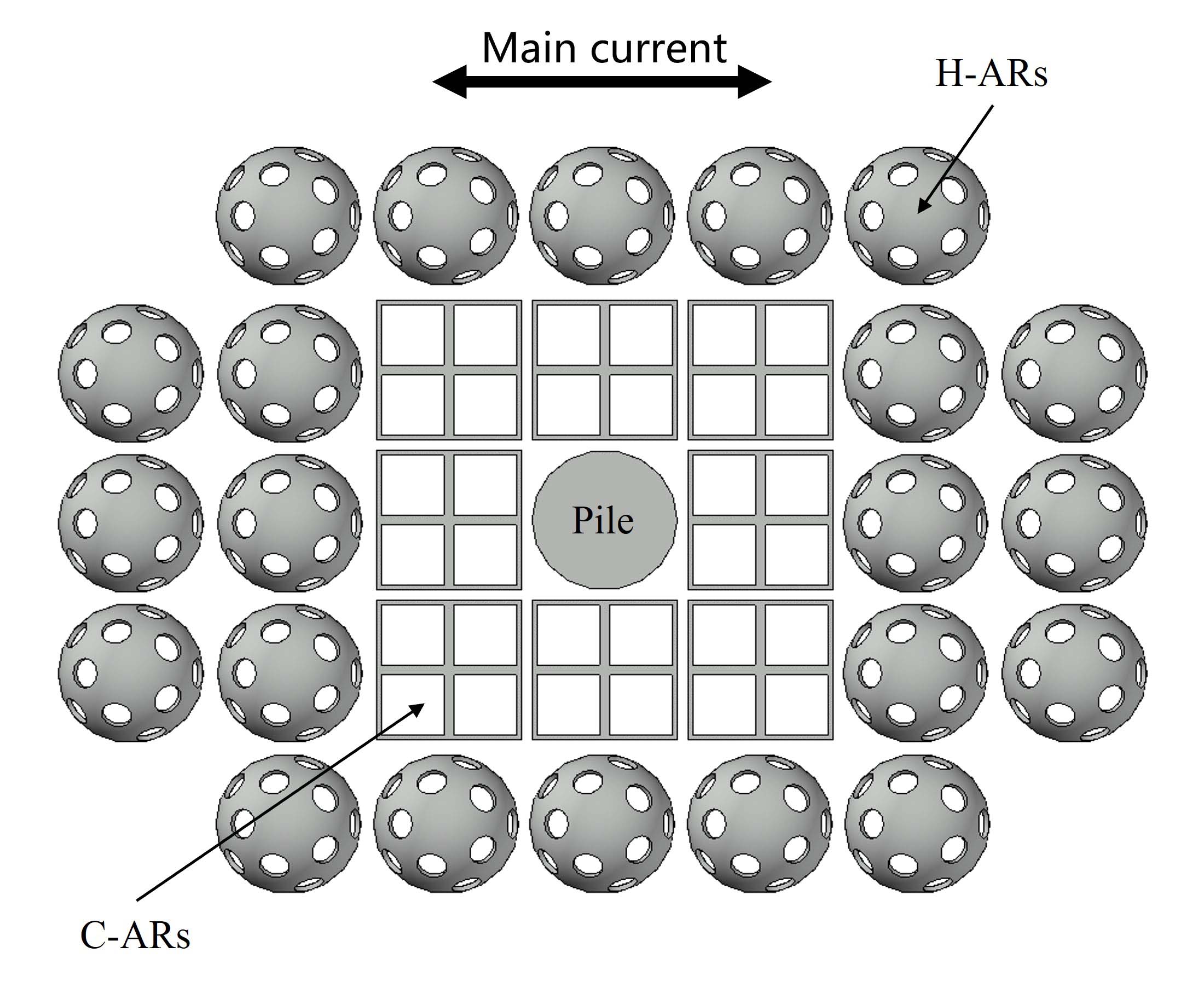}
  \caption{A suggested design of AR layout around a monopile foundation.}
  \label{fig_16}
\end{figure}

Based on the findings from our model tests, an optimized layout of AR units around a monopile is sketched in Fig.~\ref{fig_16}. Most currents in shallow coastal regions are tidal currents, which reverse their direction periodically, so the ARs should be arranged symmetrically around the monopile, and more rows of ARs should be deployed along the principle axis of tidal flow (or the main-current axis). The general concept is to place ARs that can best block the incoming flow (e.g., the C-ARs) around the pile to form a kernel, and then add some streamlined ARs that are adaptive to local scour (e.g., H-ARs) outside the kernel as buffers. In the proposed design, a layer of 8 C-ARs is deployed around the monopile, since they can best shield the monopile from the outside flow and reduce scour. Next, a layer of 16 H-ARs is deployed outside the kernel of C-ARs. These H-ARs will reduce the edge scour of C-ARs and hence limit the displacement of C-ARs. In the main-current direction, a third layer of H-ARs (3 units on the upstream side and 3 units on the downstream side) is placed to further reduce the incoming flow. The C-ARs' edge length and H-AR's diameter are both equal to (or close to) the diameter of the pile, and the ARs are tightly placed together. It is worthwhile to mention that a combination of different AR types offers extra ecological benefits by fulfilling a wider range of biological functions.

\section{Conclusion}

In this study, fixed-bed and live-bed tests under steady flow conditions are conducted to investigate the scour protection efficiency of ARs around a monopile. In fixed-bed tests, flow field information was obtained via PIV measurements, revealing how ARs change the flow around the monopile. Based on these results, several hypotheses regarding scour protection effects of ARs were put forward. Subsequently, live-bed tests are performed to verify these hypotheses and check the efficiency of some modified designs. The main conclusions of this study are summarized as follows.

When arranged around a monopile in compact 3$\times$3 pattern, both C- and H-ARs can reduce downstream flow velocity and unsteadiness by roughly 50$\sim$80\%. C-ARs can effectively eliminate the downward flow in front of the pile, while H-ARs enhance it. In the gaps between front row units of C-ARs, high-speed flow and intense unsteadiness occur due to the 'narrow gap' effect and sharp edges of the C-ARs. Additionally, C-ARs generate extensive lateral flow, which was not observed with H-ARs.

Scour results from live-bed tests can be largely explained by the flow field observed in fixed-bed tests. Near-bed areas with greater flow velocity and unsteadiness are prone to suffer severe scour. Although C-ARs can largely prevent the upstream and downstream scour, their edge scour can break down their tightly packed layout and reduce their overall scour protection. The H-ARs offer less protection against scour but are also less prone to the side effects of edge scour. Besides, the intensity of lateral flow around the AR array influences the width of the scour pit. 

Spacing ARs apart to mitigate the 'narrow-gap' effects does not compensate for the increase of unprotected areas, making it a less effective design. On the other hand, adding multiple layers of ARs in the upstream significantly improves the scour protection efficiency, e.g., further reducing scour by up to 79\%. Finally, using a combination of two AR types not only provides better scour protection but also minimizes edge scour. Based on these findings, a recommended design includes a core layer of 8 C-ARs surrounded by 22 H-ARs as outer buffers is proposed.

The research outcomes in this study are restricted by the limited test conditions, and therefore are only applicable to some simple scenarios. Some additional experiments with larger model scales and more complex flow conditions are necessary for better guiding practical engineering applications. Additionally, to establish a quantitative relationship between the required number of AR layers and specific flow conditions, more live-bed tests should be conducted. Moreover, our study focused on two typical AR types, further research can explore other AR designs, such as perforated pipes or triangular shapes. Optimizing AR structures (e.g., adjusting the size and number of openings or modifying the internal space) also remains a critical topic for future investigation.

\section*{Acknowledgments}
This research is supported by the National Key R\&D Program of China (2022YFB4201300), Science and Technology Program of China Huaneng Group Co., Ltd. (HNKJ23-H21).




\bibliographystyle{elsarticle-harv}
\bibliography{refs}





\end{document}